\documentclass[useAMS,usenatbib,onecolumn,usegraphicx]{mn2e}
\usepackage{times}
 \def\ProDiMo{{\sc ProDiMo\ }}
\newcommand{\hh}[0]{H$_2$}
\newcommand{\water}[0]{H$_2$O}
\newcommand{\UV}[0]{h$\nu$}
\newcommand{\ra}[0]{$\rightarrow$}

\title[High HDO/H$_2$O in warm dense gas]{Warm non-equilibrium gas phase chemistry as a possible origin of high HDO/H$_{\rm 2}$O ratios in hot and dense gases: application to inner protoplanetary discs}
\author[W.-F. Thi et al.]{W.-F. Thi$^{1}$ \thanks{E-mail: wfdt@roe.ac.uk}, P. Woitke$^2$, I. Kamp$^3$\\
$^1$\thanks{Scottish Universities Physics Alliances}SUPA, Institute for Astronomy, University of Edinburgh, Royal Observatory, Blackford Hill, Edinburgh, EH9 3HJ, UK\\
$^2$UK Astronomy Technology Centre, Royal Observatory, Edinburgh, Blackford Hill, Edinburgh EH9 3HJ, UK\\
$^3$Kapteyn Astronomical Institute, Postbus 800, 9700 AV Groningen, The Netherlands\\
}
  
\begin{document}

\date{Accepted 2009. Received 2009 February 12}
   
\pagerange{\pageref{firstpage}--\pageref{lastpage}} \pubyear{2009}

\maketitle

\label{firstpage}

\begin{abstract}
  The origin of Earth oceans is controversial. Earth could have
  acquired its water either from hydrated silicates (wet Earth
  scenario) or from comets (dry Earth scenario). [HDO]/[H$_2$O] ratios
  are used to discriminate between the scenarios. High [HDO]/[H$_2$O]
  ratios are found in Earth oceans. These high ratios are often
  attributed to the release of deuterium enriched cometary water ice,
  which was formed at low gas and dust temperatures. Observations do
  not show high [HDO]/[H$_2$O] in interstellar ices. We investigate
  the possible formation of high [HDO]/[H$_2$O] ratios in dense
  ($n_{\mathrm H}>$10$^6$ cm$^{-3}$) and warm gas ($T=$100--1000~K) by
  gas-phase photochemistry in the absence of grain surface
  chemistry. We derive analytical solutions, taking into account the
  major neutral-neutral reactions for gases at $T>100$~K. The chemical
  network is dominated by photodissociation and neutral-neutral
  reactions.  Despite the high gas temperature, deuterium
  fractionation occurs because of the difference in activation energy
  between deuteration enrichment and the back reactions. The
  analytical solutions were confirmed by the time-dependent chemical
  results in a 10$^{-3}$ M$_\odot$ disc around a typical TTauri star
  using the photochemical code \ProDiMo. The \ProDiMo code includes
  frequency-dependent 2D dust-continuum radiative transfer, detailed
  non-LTE gas heating and cooling, and hydrostatic calculation of the
  disc structure.  Both analytical and time-dependent models predict
  high [HDO]/[H$_2$O] ratios in the terrestrial planet forming region
  ($<$ 3~AU) of circumstellar discs. Therefore the [HDO]/[H$_2$O]
  ratio may not be an unique criterion to discriminate between the
  different origins of water on Earth.
\end{abstract}



\section{Introduction}

The water deuteration abundance ratio [HDO]/[H$_2$O] is often used to
determine the temperature at which water was synthesised. Current
sub-millimeter observations already provide measurements of
[HDO]/[H$_2$O] for many astronomical objects including comets,
hot-cores, and protoplanetary discs. In particular, a high
[HDO]/[H$_2$O] has been observed toward the protoplanetary disc DM~Tau
\citep{Ceccarelli2005ApJ...631L..81C}, although the HDO detection
remains controversial \citep{Guilloteau2006A&A...448L...5G}. The
Herschel Space Telescope will be capable to detect cool and warm water
in discs while the Atacama Large Millimeter Array (ALMA) will
spatially resolve HDO emission.

The water deuteration abundance ratio [HDO]/[H$_2$O] in hot cores
around massive young stellar objects is $\sim 3 \times 10^{-4}$
\citep{Gensheimer1996A&A...314..281G}, around a factor $\sim$ 10
enhancement compared to the cosmic [D]/[H] value set at the Big Bang
($\sim$~1.5$\times$10$^{-5}$, \citealt{Linsky2003SSRv..106...49L}).
The enrichment suggests that the chemistry occurs in a low density,
cold medium ($T<$~100~K), where HDO synthesis begins with the
deuteration of the molecular ion H$_3^+$ into H$_2$D$^+$ (e.g.,
\citealt{Roberts2000A&A...361..388R}). Further gas-phase reactions
lead to a high [HDO]/[H$_2$O]. But at temperatures greater than 100~K,
which is the typical gas temperature of hot cores, the initial
deuteration reaction is inefficient and the [HDO]/[H$_2$O] is close to
the cosmic value. Therefore current chemical models need to invoke the
evaporation of deuterium enriched water ice when the temperature
reaches 100~K to explain the deuterium enrichment in water. The
observed (HDO/H$_2$O)ice ratio is however too low
\citep{Dartois2003A&A...399.1009D,Parise2003A&A...410..897P}. A second
difficulty arises from the high observed water ice abundance of
10$^{-5}$--10$^{-4}$, which is two orders of magnitude larger than the
gas phase abundance of water in the hot core around IRAS 16293-2422
\citep{Parise2005A&A...431..547P}. 

After decades of research, the origin of water on Earth remains a
major subject of debate (e.g., \citealt{Nuth2008EM&P..102..435N} for a
review).  Two majors models exist. In the first model, the Earth
accretes dry: the building blocks of the Earth are made of dry rocks,
composed only of silicates and carbonaceous materials. Most of the water
is brought afterward by comets during the phase of heavy
bombardment. This model is supported by the similarity between the
Earth Mean Ocean Water (SMOW $\simeq$~1.49$\times$10$^{-4}$)
[HDO]/[H$_2$O] and the cometary value, although the later value seems
too high. Another support to the "dry" model is the low [HDO]/[H$_2$O]
predicted at temperature $T>$~100~K by thermochemical equilibrium models. In
the second model, most of the water on Earth comes from the release of
water vapor trapped inside the water-entrapped planetesimals like
carbonaceous chondrites upon impact or during volcanism. Water-rich
planetesimals located at 2-3~AU are perturbed by the giant planets and
collide with the young Earth
\citep{Morbidelli2000M&PS...35.1309M,Gomes2005Natur.435..466G,Raymond2004Icar..168....1R,Raymond2005ApJ...632..670R}. Those
planetesimals contain water in form of hydrated silicates.  The
average D/H in carbonaceous chrondrite is similar to the value in the
Earth ocean \citep{Robert2000SSRv...92..201R}, although the
composition of carbonaceous chrondrite does not match the composition
of Earth'crust \citep{Righter2006mess.book..803R}. The Earth is said
to have accreted "wet" \citep{Drake2005M&PS...40..519D}. Hydrous
material does not necessary need to be brought from the asteroid belt
area (2-3~AU). Recent study on adsorption of water molecules onto
fractal dust grains shows that hydrous material could have been
present in the vicinity of 1~AU \citep{Stimpfl2006JCrGr.294...83S}. In
this scenario water in the gaseous solar nebula sticks onto the
silicate surface of fractal grains, which coagulate and grow into
the Earth. The adsorbed water would reflect the deuterium enrichment
of the gas-phase water at 1~AU. The major difference between the two
``wet'' scenarios is that water was present in the early-Earth in the
latter scenario. Geochemical findings suggest that hydrosphere and
continental crust were already present in the first $\sim$~600 million
years of Earth history \citep{Hopkins2008Nature}. The presence of
water and also of organic matter at the formation of the Earth has
profound astrobiology implications for the origin of life.

One major weakness of the ``wet'' scenario is that the deuterium
enrichment predicted by equilibrium chemistry is too low at
temperatures greater than 100~K. However there are chemical routes to
obtain high [HDO]/[H$_2$O] ratios at temperatures greater than 100~K
in non-equilibrium chemistry. Photodissociation of H$_2$O and HDO
followed by reformation at the surface of protoplanetary discs and
turbulent mixing can also change the [HDO]/[H$_2$O]
\citep{Yung1988Icar...76..146Y}. Vertical turbulent mixing was invoked
by \cite{Lyons2005Natur.435..317L} to explain the oxygen anomalies in
the early solar nebula.  Alternatively \cite{Genda2008Icar..194...42G}
propose that the [HDO]/[H$_2$O] ratio on Earth changed during the
evolution of the Earth from its
formation. \citet{Willacy2009ApJ...703..479W} studied the deuterium
chemistry of the inner disc around a TTauri star and found relatively
high HDO abundance in the warm molecular layer.

Four direct physical processes lead to isotopic fractionation in the
gas phase. First, heavier isotopologues diffuse slower than the
lighter main isotopologue. The diffusion results in zones of varying
[HDO]/[H$_2$O] ratios. This process is negligible in turbulent media
when the mixing timescale is much lower than the diffusion
timescales.  Second, due to their larger reduced mass, isotopologues
usually have a lower zero point energy. The vibrational energy levels
are located lower, which increases the density sum of energy and hence
reduces the vapor pressure. Hence, the water ice/gas ratios in the
interstellar medium is not at equilibrium. Third, heavier
isotopologues are chemically more stable because the binding energy
between O and D is stronger than between O and H in the water
molecule. Fractionation reactions dominate at low temperature because
the inverse reactions have a minimum energy barrier equal to the
difference in binding energy.  Finally, isotopologues differ in their
photodissociation cross section due to changes in the selection
rules. A combination of the last two processes together with chemical
diffusion and turbulent mixing probably determine the deuterium
fractionation in the inner region of protoplanetary discs where
terrestrial planets form.

Interstellar chemical models often neglect or include only few
neutral-neutral reactions because their rates are small at temperatures
below $\sim$~100~K but they become competitively fast with the
ion-neutral reactions at a few hundred degrees. For example, in dense
and warm gases, the central species is the hydroxyl radical OH, which
can react with H$_2$ to form H$_2$O \citep{Thi2005A&A...438..557T}.

In this paper, we explore neutral-neutral and photochemical reactions
relevant to the formation and destruction of H$_2$, HD, OH, OH, H$_2$O
and HDO at gas temperature between 100 and 1000~K. We first focus on
steady-state abundances and perform an analytical analysis. Analytical
analysis makes it possible to determine the main formation and
destruction paths. After establishing the possible reaction paths, we
use the photochemical code \ProDiMo to compute a time-dependent
chemical structure for a typical TTauri disk to check the validity of
our assumptions in the analytical analysis. In the rest of the paper,
the chemical reaction rates are given in
Table~\ref{tab_PrincipleReac}, \ref{tab_Reactions1},
\ref{tab_Reactions2}, \ref{tab_Reactions3}, and
\ref{tab_Reactions4}. In particular, the most important reactions are
summarised in Tables~\ref{tab_PrincipleReac}.


\section{Steady-state analytical solution}\label{network}

\subsection{Formation and destruction of H$_2$O and HDO}

We consider a gas mixture at a single temperature, density, irradiated
by UV photons. In thermochemical equilibrium the deuterium exchange
equilibrium reaction between water and molecular hydrogen (the most
abundant reservoir of deuterium) is (e.g.,
\citealt{Robert2000SSRv...92..201R})
\begin{equation}
HD + H_2O \leftrightarrow HDO + H_2.
\end{equation}
The equilibrium constant of this reaction for gas between 273 and
1000~K is \citep{Richet1977AREPS...5...65R}
\begin{equation}
K=\frac{P_{HDO}P_{H_2}}{P_{HD}P_{H_2O}}\simeq \frac{0.22\times10^6}{T^2}+1.
\end{equation}
The pressure dependence of $K$ is negligible. At equilibrium, the
deuterium enrichment decreases quadratically with temperature. A gas
mixture is in thermochemical equilibrium when each chemical reaction is
exactly balanced by its reverse reaction. Dissociations upon
absorption of UV photons are not taken into account.

A gas is in chemical steady-state (also called stationary state) if
the various abundances do not vary with time but the formation and
destruction reactions for a given species are not necessarily the
same. An analytical solution to the chemical network for the synthesis
of H$_2$O and HDO can be obtained by identifying the main formation
and destruction chemical reactions.  At high temperature ($T>$100~K)
and density, it is well established that water is primarily formed via
the reactions between the hydroxyl radical OH and molecular hydrogen
H$_2$ (e.g,
\citealt{Thi2005A&A...438..557T,Glassgold2009ApJ...701..142G}). High
temperature gases are needed to overcome the large energy barrier of
the reaction (reaction 5 with rate $k_5$, $E_a$=1660~K). The main
destruction mechanism is likely photodissociation in low
A$_\mathrm{V}$ regions. Other water destruction mechanisms are
reactions with atomic hydrogen, and with ionised helium in shielded
regions. We assume in this study that the dust temperature
$T_{\mathrm{d}}$ has equilibrated with the gas temperature. At
$T_{\mathrm{d}}>$100~K most water molecules remain in the gas
phase. The rate of water formation is thus:
\begin{equation}
\frac{d[H_2O]}{dt}=(k_5[H_2]+k_{18}[HD]+k_{7}[OH])[OH]-(J_6+k_8[H]+k_{19}[D]+k_{41}[He^+]+k_{cp,3})[H_2O],
\end{equation}
where $k_5$, $k_{18}$, and $k_{7}$ are the rates of formation of water via the reactions OH+H$_2$, OH+HD, and OH+OH respectively, $J_6$ is the photodissociation rate, $k_8$ is the water destruction
rate by H, $k_{cp,3}$ is the cosmic-ray induced photodissociation of
water, and $k_{41}$ is the rate of destruction by He$^+$. At steady-state ($d[H_2O]/dt=0$), 
we obtain the ratio
\begin{equation}
\frac{[H_2O]}{[OH]}\simeq\frac{k_5[H_2]}{J_6+k_8[H]+k_{41}[He^+]+k_{cp,3}},
\label{eq_H2O_OH_ratio}
\end{equation}
which will be used later in this paper. We have neglected the
formation of water via reaction of OH with HD and OH and destruction
of water via atomic deuterium because the rates are orders of
magnitude smaller than the rates of other paths. Deuterated water can
be formed via the reaction of HD with OH as well as via OD + H$_2$
\citep{Bergin1998ApJ...499..777B}. Destruction occurs via
photodissociation, reaction with atomic hydrogen and ionised helium.
\begin{equation}
\frac{d[HDO]}{dt}=k_{22}[OD][H_2]+k_{20}[OH][HD]-(J_{7a}+J_{7b}+(k_{21}+k_{23})[H]+(k_{42}+k_{43})[He^+]+k_{cp,4}+k_{cp,5})[HDO].
\end{equation}
\noindent
The photodissociation of HDO can either lead to OH + D ($J_{7a}$) or
to OD + H ($J_{7b}$). The sum of the two photorates is the total
HDO photodissociation rate and is similar to the H$_2$O
photodissociation rate $J_6$. The photodissociation of HDO favors
cleavage of the O-H bond over O-D bond with a 3 to 1 ratio
\citep{Shafer1989,VanderWall1990,VanderWall1991}, i.e. $J_{7b} \simeq 3 J_{7a}$
because the binding energy between O and D is stronger than that
between O and H in water. In steady-state, the balance between
formation and destruction leads to the ratio
\begin{equation}
X_D=\frac{[HDO]}{[H_2O]}=\frac{(k_{22}[OD][H_2]+k_{20}[OH][HD])/(J_{7a}+J_{7b}+(k_{21}+k_{23})[H]+(k_{42}+k_{43})[He^+]+k_{cp,4}+k_{cp,5})}{k_5[OH][H_2]/(J_6+k_8[H]+k_{41}[He^+]+k_{cp,3})},
\end{equation}
\begin{equation}
X_D=\frac{[HDO]}{[H_2O]}=\left(\frac{D_{H_2O}}{D_{HDO}}\right) \left(\frac{k_{22}([OD]/[OH])[H_2]+k_{20}[HD]}{k_5[H_2]}\right),\label{eq_XD}
\end{equation}
where we define
\begin{equation}
D_{H_2O}=J_6+k_8[H]+k_{41}[He^+]+k_{cp,3}
\end{equation}
and
\begin{equation}
D_{HDO}=J_{7a}+J_{7b}+(k_{21}+k_{23})[H]+(k_{42}+k_{43})[He^+]+k_{cp,4}+k_{cp,5}
\end{equation}
to be the total destruction rates of H$_2$O and HDO. The total UV
photodissociation rates of H$_2$O and HDO are similar
\citep{ZhangImre1988}: $J_6 \simeq J_{7a}+J_{7b}$. We further assume
that rates with the deuterated species are similar to that of the main
isotopologues: $k_8=k_{21}+k_{23}$, $k_{41}\simeq k_{42}+k_{43}$, and
$k_{cp,4}+k_{cp,5}=k_{cp,3}$. 
If we can make the assumption that the total destruction rate of
H$_2$O and HDO are similar, i.e. $D_{H_2O}\simeq D_{HDO}$, then the value
of [HDO]/[H$_2$O] does not depend on the actual destruction
mechanisms for H$_2$O and HDO.  The water fractionation $f$(HDO) is defined as
\begin{equation}
  f(HDO)=\frac{[HDO]/[H_2O]}{[HD]/[H_2]} = \left(\frac{D_{H_2O}}{D_{HDO}}\right) \left( \frac{k_{22}}{k_5}\frac{[OD]/[OH]}{[HD]/[H_2]}+\frac{k_{20}}{k_{5}} \right)= \left(\frac{D_{H_2O}}{D_{HDO}}\right) \left(\frac{k_{22}}{k_5}f(OD)+\frac{k_{20}}{k_{5}}\right) \simeq \frac{k_{22}}{k_5}f(OD)+\frac{k_{20}}{k_{5}}
\end{equation}
The ratio $k_{20}/k_{5}$ is the ratio between the reaction of OH with
HD to form HDO compared to the reaction of OH with H$_2$ to form
water.  From the values in Table~\ref{tab_PrincipleReac} the
$k_{20}/k_{5}$ ratio is lower than 1 for gas temperature greater than
120~K. This term does not enhance the water deuteration
fractionation. We can write analytically the ratio between rate
$k_{22}$ and rate $k_5$:
\begin{equation}
\frac{k_{22}}{k_5}=\frac{1.55\times 10^{-12}(T/300)^{1.6}e^{-1663/T}}{2.05\times 10^{-12}(T/300)^{1.52}e^{-1660/T}}=0.756098(T/300)^{0.08}e^{-3/T}.
\end{equation}
The ratio is weakly temperature-dependant. Since $k_{22}/k_5 = 0.7-1$
between 100 and 1000~K, water and hydroxyl radical deuterium fraction
are similar at gas temperature greater than $\sim$~200~K:
$f$(HDO)$\simeq$~$f$(OD).  Therefore water would be deuterium
fractionated if OH is (i.e. $f$(OD)$>$1).

\subsection{Formation and destruction of OH  and OD}

The water deuteration fraction $f$(HDO) is intimately linked to that
OH $f$(OD). The rate limiting reaction for the formation of water is
the formation of hydroxyl radical OH: O + H$_2$ $\rightarrow$ OH + H
with rate $k_1$. This reaction has a large energy barrier
($E_a$=3163~K). The chemistry of OD (and OH) is described in details
by \cite{Croswell1985ApJ...289..618C} for reactions without
barrier. The production of OD mostly occurs through the rapid exchange
reaction
\begin{equation}
D + OH \rightarrow OD + H
\end{equation}
\noindent
once OH is present in the gas (reaction 16 with rate from
\citealt{Yung1988Icar...76..146Y}). The forward reaction has no
activation barrier but the reverse reaction (reaction 13) has a
barrier of 810~K because OD is more stable than OH. Thus at
$T<$1000~K, OD is favored. The role of OD for gas between 100 and
1000~K is similar to that of H$_2$D$^+$ for gas at $T<$~100~K for 
the deuterium enrichment. OD can also formed by the reaction
\begin{equation}
HD + O \rightarrow OD + H.
\end{equation}
OD is destroyed by reaction with carbon ion
\begin{equation}
C^+ + OD \rightarrow CO + D^+
\end{equation}
\begin{equation}
C^+ + OD \rightarrow CO^+ + D,
\end{equation}
\noindent
and by photodissociation (with rate $J_5$)
\begin{equation}
OD + h\nu \rightarrow O + D.
\end{equation}
The steady-state abundance of OH and OD taking into account the most important formation and destruction reactions are
\begin{equation}
[OH]=\frac{k_1[O][H_2]+(J_6+k_8[H]+k_{41}[He^+]+k_{cp,3})[H_2O]+k_{14}[O][HD]+k_{22}[OD][H_2]+k_{30}[CO][H]+(J_{7a}+k_{cp,5})[HDO]}{(k_{15}+k_{16})[D]+k_4[H]+k_5[H_2]+k_{44}[He^+]+k_6[O]+k_{26}[OD]+k_{28}[CO]+k_{29}[C]+k_{39}[C^+]+J_4+k_{cp,1}}
\end{equation}
\begin{equation}
[OH] \simeq \frac{k_1[O][H_2]}{k_4[H]+k_5[H_2]+k_{28}[CO]+J_4+k_{cp,1}+k_{29}[C]+k_{39}[C^+]}
\end{equation}
After some algebra, we obtain the ratio
\begin{equation}
\frac{[OH]}{[O]} \simeq \frac{k_1[H_2]}{J_4+k_4[H]+k_{28}[CO]+k_{cp,1}+k_{29}[C]+k_{39}[C^+]+k_{44}[He^+]+k_{cp,1}},
\end{equation}
which would be useful later in the analysis. For the deuterated
hydroxyl radical at steady-state:
\begin{equation}
[OD]=\frac{k_{16}[OH][D]+k_{12}[O][HD]+k_{23}[HDO][H]+k_{25}[O_2][D]+k_{27}[HDO][O]+k_{33}[CO][D]+(J_{7b}+k_{cp,4})[HDO]}{k_{17}[H]+k_{22}[H_2]+k_{45}[He^+]+k_{24}[O]+k_{26}[OH]+k_{31}[CO]+k_{32}[C]+k_{40}[C^+]+J_5+k_{cp,2}}.
\end{equation}

\noindent
The rates are listed in Table 1 to 5. Neglecting the minor formation and destruction reactions (i.e. reactions with C and C$^+$), we simplify the ratio:

\begin{equation}
\frac{[OD]}{[OH]}\simeq \frac{D_{OH}}{D_{OD}}\left(\frac{k_{16}[OH][D]+k_{12}[O][HD]+(J_{7b}+k_{23}[H]+k_{cp,4})[HDO]}{k_1[O][H_2]+(J_6+k_8[H]+k_{cp,3})[H_2O]} \right),\label{eq_OD_OH_ratio}
\end{equation}
where 
\begin{equation}
\frac{D_{OH}}{D_{OD}}=\frac{(k_{15}+k_{16})[D]+k_4[H]+k_5[H_2]+k_{44}[He^+]+k_6[O]+k_{26}[OD]+k_{28}[CO]+k_{29}[C]+k_{39}[C^+]+J_4+k_{cp,1}}{k_{17}[H]+k_{22}[H_2]+k_{45}[He^+]+k_{24}[O]+k_{26}[OH]+k_{31}[CO]+k_{32}[C]+k_{40}[C^+]+J_5+k_{cp,2}}
\end{equation}
is the ratio between the sum of all destruction rates of OH and that
of OD. The photodissociation rate of OH (rate $J_4$) and OD (rate
$J_5$) are close
\citep{vanDishoeck1984ApJ...277..576V,vanDishoeck1988rcia.conf...49V,Croswell1985ApJ...289..618C}
($J_4 \simeq J_5$).  From Table~\ref{tab_PrincipleReac} reactions 5
and 22 have similar rates ($k_5 \simeq k_{22}$). Reactions 52 and 53
are ion-molecule reactions and we expect the rates to be of the same
order of magnitude.  Only reactions 4 (OH + H \ra O + H$_2$) and 17
(OD + H \ra OH + D) have very different rates, reaction 17 being much
faster. But those reactions are important at low A$_\mathrm{V}$ only
where photodissociation dominates as destruction process. We assume
that $k_6\simeq k_{24}$.

The hydroxyl radical deuterium fractionation reads

\begin{equation}
f(OD)=\frac{[OD]/[OH]}{[HD]/[H_2]}=\frac{D_{OH}}{D_{OD}}
\left(\frac{k_{16}([OH]/[O])[D]+k_{12}[HD]+(J_{7b}+k_{23}[H]+k_{cp,4})X_D([H_2O]/[O])}{k_1[H_2]+(J_6+k_8[H]+k_{cp,3})([H_2O]/[O])}\right)/\left(\frac{[HD]}{[H_2]}\right).
\end{equation}

The OD fractionation increases with larger amount of OH and atomic
deuterium and decreases if oxygen is in the atomic form and deuterium
is locked in HD. Large amount of OH allows the deuterium exchange
reaction to occur. 

The knowledge of the abundances of H, H$_2$, D, and HD is needed to estimate
the hydroxyl deuterium fraction $f(OD)$.

\subsection{H and H$_2$}

The formation of H$_2$ occurs mostly on grain surface when the grain
temperature is below 1000~K and the destruction is caused by
photodissociation at low A$_V$ and cosmic-ray/X-ray induced ionization
and reaction with He$^+$ in the UV free region. The temperature is low
enough to avoid H$_2$ destruction by atomic hydrogen. The steady-state
balance between formation and destruction is:
\begin{equation}
R_1(T_g,T_d)n_H[H]=(f_{ss}J_1+k_{\zeta,4}+k_{35}[He^+])[H_2],	
\end{equation}
where $f_{ss}$ is the H$_2$ self-shielding function \citep{Draine1996ApJ...468..269D} and $R_1(T_g,T_d)$ the molecular hydrogen formation rate on grain surfaces \citep{Tielens2005pcim.book.....T}
\begin{equation}
R_1(T_g,T_d)=4.4\times 10^{-17} S(T_g,T_d)\left(\frac{T_g}{100}\right)^{1/2},
\end{equation}
where $T_g$ and $T_d$ are the gas and dust grain temperature respectively and the sticking coefficient $S$ is defined as
\begin{equation}
S(T_g,T_d)=\frac{1}{1+4\times 10^{-2}(T_g+T_d)^{1/2}+2\times 10^{-3}T_g+8\times 10^{-6}T_g^2}.
\end{equation}
The sticking coefficient ensures that at high dust temperature atomic hydrogen does not stick onto grain surfaces. The number density of nuclei is
\begin{equation}
n_H=[H]+2[H_2]	
\end{equation}
We obtain the atomic and molecular abundances:
\begin{equation}
[H] = n_H\left(\frac{f_{ss}J_1+k_{\zeta,4}+k_{35}[He^+]}{2R_1n_H+f_{ss}J_1+k_{\zeta,4}+k_{35}[He^+]}\right)\simeq n_H\left(\frac{f_{ss}J_1+k_{\zeta,4}}{2R_1n_H+f_{ss}J_1+k_{\zeta,4}}\right)
\end{equation}
and
\begin{equation}
[H_2]=n_H\left(\frac{R_1n_H}{2R_1n_H+f_{ss}J_1+k_{\zeta,4}+k_{35}[He^+]}\right)\simeq n_H\left(\frac{R_1n_H}{2R_1n_H+f_{ss}J_1+k_{\zeta,4}}\right).
\end{equation}
Atomic and molecular hydrogen abundances are determined by the gas
temperature, density, UV flux, extinction, the self-shielding
function, and cosmic ray flux.

\subsection{D and HD}

While H$_2$ is predominately formed on grain surfaces, formation of HD
can also occur in the gas phase by reaction between D and H$_2$ at
high temperature and at density higher than about 5 $\times$ 10$^3$
cm$^{-3}$ \citep{LePetit2002A&A...390..369L}. HD is destroyed by
photodissociation and reaction with atomic H while H$_2$ is mainly
photodissociated. H$_2$ can self-shield against photodissociation, HD
is shielded by dust only. Therefore deuterium remains atomic at higher
extinction than H$_2$. In hot and dense regions, an important
formation route of HD is therefore: H$_2$ + D \ra\ HD + H (reaction 10
with a barrier $E_a$=3820~K).
\noindent
Larger amount of atomic deuterium is needed to obtain high OD over OH
ratio.  The production of atomic deuterium is enhanced at high
temperature via the conversion reaction HD + H \ra\ H$_2$ + D. Another
destruction mechanism of HD molecules involves photodissociation: HD +
\UV\ \ra\ H + D. Finally, reaction with H exchanges the atomic
hydrogen with atomic deuterium: HD + H \ra\ H$_2$ + D. At high
A$_\mathrm{V}$, cosmic rays destroy HD. Reactions with H$^+$ and
atomic oxygen also destroy HD. The steady-state balance for [HD] then
reads

\begin{equation}
[HD]=\frac{(k_{10}[H_2]+R_{2}n_H)[D]+k_{37}[D^+][H_2]}{J_2+k_{\zeta,5}+k_{\zeta,6}+k_{11}[H]+k_{36}[H^+]+(k_{12}+k_{14})[O]+k_{18}[OH]},
\end{equation}
\noindent where $R_2$ is the formation rate on grain surfaces
\citep{LePetit2002A&A...390..369L} 
density
\begin{equation}
R_2(T_g,T_d)=6.3\times 10^{-17} S(T_g,T_d)\left(\frac{T_g}{100}\right)^{1/2}
\end{equation}
and $n_H$ is the total number. This H$_2$ formation rate does not take
into account grain chemisorption sites contrary to more sophisticated
H$_2$ formation model \citep{Cazaux2002ApJ...575L..29C}. We assume
$k_{\zeta,5}+k_{\zeta,6} \simeq k_{\zeta,4}$ and that the sticking
coefficient is the same than for H$_2$.

We neglect the formation of HD via reaction with D$^+$ and destruction
via reactions with the protons, atomic oxygen and OH radicals.  The
steady-state balance leads to
\begin{equation}
\frac{[D]}{[HD]}=\frac{J_2+k_{\zeta,4}+k_{11}[H]}{k_{10}[H_2]+R_2n_H}
\end{equation}
If we can assume that most of the deuterium is locked in H and HD
\begin{equation}
n_D \simeq [D] + [HD],
\end{equation}
then we obtain HD and D abundances
\begin{equation}
[HD] = n_D\left(\frac{k_{10}[H_2]+R_2n_H}{J_2+k_{\zeta,4}+k_{11}[H]+k_{10}[H_2]+R_2n_H}\right)
\end{equation}
and
\begin{equation}
[D] = n_D
\left(\frac{J_2+k_{\zeta,4}+k_{11}[H]}
{J_2+k_{\zeta,4}+k_{11}[H]+k_{10}[H_2]+R_2n_H}\right).
\end{equation}
At low A$_{\mathrm{V}}$, destruction of [HD] is dominated by photodissociation and at gas temperature a few 100~K, the formation of HD via reaction of atomic deuterium with molecular hydrogen is negligible 
\begin{equation}
\frac{[D]}{[HD]}\simeq\frac{J_2}{R_2n_H}.
\end{equation}
Even in obscured ($J_2 \sim 0$) and fully molecular regions ($[H] \sim
0$), cosmic-rays induced photodissociation and deuterium exchange reactions ensure that some atomic deuterium always remains in atomic form:
\begin{equation}
\frac{[D]}{[HD]}\simeq\frac{k_{\zeta,4}}{R_2n_H}.
\end{equation}
The formation of HD on grains decreases dramatically for dust grain temperature above 100~K.

\subsection{Neutral and ionised Helium}

Where the steady-state abundance of ionised Helium  is given by the expression:
\begin{equation}
[He^+]=\frac{k_{\zeta,3}[He]}{k_{35}[H_2]+k_{e^-,3}[e^-]}
\end{equation}
\noindent Combining with the element conservation equation:
\begin{equation}
n_{He}=[He]+[He^+]
\end{equation}
\noindent we obtain
\begin{equation}
[He^+]=\frac{k_{\zeta,3}n_{He}}{k_{35}[H_2]+k_{e^-,3}[e^-]-k_{\zeta,3}}
\end{equation}
The He$^+$ abundance can be estimated only if the electron abundance
is known. Our simple analytical analysis cannot provide an estimate of
the He$^+$ abundance and we will omit in the rest of the paper
reactions with He$^+$.
\subsection{Water fractionation ratio $f(HDO)$}

The results of the previous sections can be combined to derive an analytical formula
of the [HDO]/[H$_2$O] ratio:
\begin{equation}
X_D=\frac{[HDO]}{[H_2O]}=num/den,
\end{equation}
where
 \begin{equation}
num = \frac{k_{22}}{k_5}\left(\frac{D_{OH}}{D_{OD}}\frac{D_{H_2O}}{D_{HDO}}\right)\frac{k_{16}([OH]/[O])[D]+k_{12}[HD]}{k_1[H_2]+(J_6+k_8[H]+k_{cp,3})([H_2O]/[O])}+\frac{k_{20}}{k_5}\frac{[HD]}{[H_2]}
\end{equation}
and
\begin{equation}
den = 1 - \frac{k_{22}}{k_5}\left(\frac{D_{OH}}{D_{OD}}\frac{D_{H_2O}}{D_{HDO}}\right)\frac{(J_{7b}+k_{23}[H]+k_{cp,4})([H_2O]/[O])}{k_1[H_2]+(J_6+k_8[H]+k_{cp,3})([H_2O]/[O])}
\end{equation}
The abundances and abundance ratios in $num$ and $den$ have been
derived earlier:
\begin{equation}
\frac{[OH]}{[O]} \simeq \frac{k_1[H_2]}{J_4+k_4[H]+k_{28}[CO]+k_{29}[C]+k_{39}[C^+]+k_{44}[He^+]+k_{cp,1}},
\end{equation}
\begin{equation}
\frac{[H_2O]}{[OH]}=\frac{k_5[H_2]}{J_6+k_8[H]+k_{41}[He^+]+k_{cp,3}},
\end{equation}
and combining the two equations above, we obtain
\begin{equation}
\frac{[H_2O]}{[O]}=\frac{k_1k_5[H_2]^2}{(J_4+k_4[H]+k_{28}[CO]+k_{29}[C]+k_{39}[C^+]+k_{44}[He^+]+k_{cp,1})(J_6+k_8[H]+k_{41}[He^+]+k_{cp,3})}.
\end{equation}
The water fractionation is composed of 3 terms. At gas temperature
greater than 200~K the second term $k_{12}/k_1$ dominates for both low
and high extinction and the fractionation reaches 10-1000.  Our
analysis concerns high temperature gas only.  In warm and dense
region, the main oxygen carrier is water and that of carbon is methane
CH$_4$ and not CO at high A$_{\mathrm{V}}$ and C$^+$ at low
A$_{\mathrm{V}}$. As state before, we further neglect reactions with
He$^+$.

At temperature below 100~K, the only fast chemical reactions are
between ions and neutral species, which are not taken into account in
our analysis. The starting point of water formation is H$_3^+$.
Deuterium enrichment occurs via the deuterated equivalent of H$_3^+$,
H$_2$D$^+$. H$_2$D$^+$ has a lower zero-point energy, which favors the
deuteration reaction H$_3^+$ + HD \ra H$_2$D$^+$ + H$_2$. The rate of
the back reaction has an energy barrier of 230~K. 

\begin{figure}
\centering
\resizebox{\hsize}{!}
  {\includegraphics[angle=90]{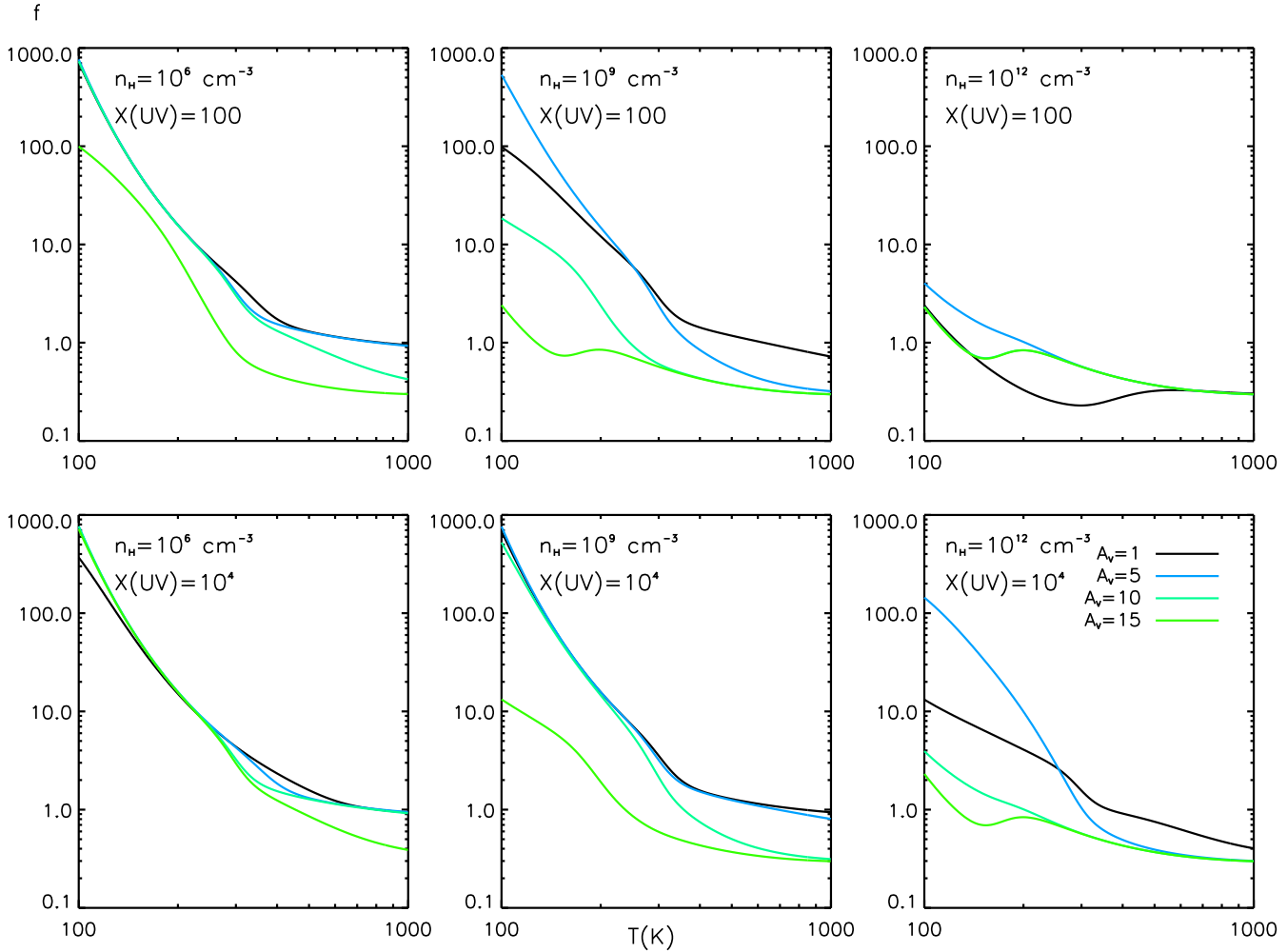}}
  \caption{Water deuteration fraction $f$(HDO). The upper panels show
    models with UV enhancement of factor 100 w.r.t. the standard
    interstellar UV filed and for three increasing gas densities. The
    four curves correspond to the various dust extinctions. The lower
    panels are models with UV enhancement of 10$^{4}$.}
  \label{fig_fractionation}
\end{figure}

The water deuteration fraction enhancement is caused by the fast
deuterium exchange reaction with OH (D+OH \ra OD + H). The inverse
reaction has an energy barrier of 717~K.  Significant amount of atomic
deuterium is possible at low and intermediate extinction. We plotted
in Fig.~\ref{fig_fractionation} the ratio [HDO]/[H$_2$O] as function
of the gas temperature for an impinging UV enhanced by a factor 10$^4$
and 10$^2$, for different extinctions (A$_{\mathrm{V}}$=1, 5, 10, 15),
and for two gas densities ($n_{\mathrm{H}}=$~10$^{8}$ and 10$^{12}$
cm$^{-3}$). Disc surfaces around young accreting T~Tauri stars receive
$\sim$ 10$^4$ to 10$^6$ times the amount of standard interstellar
UV. The UV field, extinction, density, and gas temperature values are
typically found in regions of discs where water is abundant. In the
following section, we will use a time-dependent thermo-photochemical
code to support our assumed values for the parameters.  The figure
shows that large $f$(HDO)=[HDO]/[H$_2$O] ratios (up to a few 100) are
possible for gas temperatures up to 500-600~K and up to extinction of
10, although the average value for $f$ is lower than 10. At $T>$500,
the back reaction OD+H \ra OH+D starts to destroy efficiently OD.  The
figure also shows the large range of values for $f$(HDO)
(0.1--10$^3$). The $f$(HDO) curves testify of the sensitivity of the
abundances on the gas temperature, which reflects the exponential
nature of the Arrhenius' law for neutral-neutral reactions.

\section{Time-dependent modelling}

\subsection{\ProDiMo photochemical code}

We modelled the chemical abundances of a 10$^{-3}$ M$_{\odot}$ disc
around a typical TTauri star ($T_{\mathrm{eff}}$~4400~K, $\log g$=4.0,
$Z$=1.0) using the photochemical code \ProDiMo. \ProDiMo combines
frequency-dependent 2D dust-continuum radiative transfer, kinetic
gas-phase and UV photochemistry, ice formation, and detailed non-LTE
heating and cooling balance. The major improvement over previous
studies is that the density structure is determined by the gas
pressure that is computed by detailed chemistry and energy balanced
and not by assuming that the gas and dust have the same
temperature. Detailed description of the code are given by
\citet{Woitke2009A&A...501..383W} and in \citet{Kamp2009}. The code
has been used to determine the water abundance in the inner disc
around a typical Herbig~Ae star \citep{Woitke2009A&A...501L...5W}. The
most recent additions to the code includes PAH chemistry,
time-dependent chemistry, deuterium chemistry, and generation of
Spectral Energy Distribution and spectral lines. The last two features
are not used in the modelling performed for this paper.
Table~\ref{tab_DiscParameters} summarized the input parameters to the
model. The stellar spectrum was generated using {{\sc Phoenix}}
\citep{Brott2005ESASP.576..565B} with the addition of chromospheric
flux from HD~129333 \citep{Dorren1994ApJ...428..805D}. Although the
model extends to 300~AU, we focus here only on the inner 3~AU as we
are interested in the [HDO]/[H$_2$O] ratio in the terrestrial planet
forming region of discs.

The chemical network includes a total of 187 deuterated and
non-deuterated gas and ice species. Most reaction rates are taken from
the {\sc UMIST database} \citep{Wooddall2007A&A...466.1197W}. Additional
reaction rates were compiled from the {\sc NIST} chemical kinetic
database. The rates involving deuterated species are described in
e.g., \citet{Roberts2004A&A...424..905R},
\citet{Roberts2000A&A...361..388R},
\citet{Charnley1997ApJ...482L.203C}, and
\citet{Brown1989MNRAS.237..661B}.  Species can freeze-out onto grain
surfaces and desorb thermally or upon absorption of a cosmic-ray or a
UV photon (photodesorption). Grain surface reactions were omitted
apart from the grain surface formation of H$_2$ and HD
\citep{Cazaux2002ApJ...575L..29C}.  The photodissociation
cross-sections are taken from the {\sc Leiden database} described in
\citet{vanDishoeck2008}.

The chemical abundances in the disc were established in three stages.
First \ProDiMo determined the UV field, hydrostatic density, gas and
dust temperature, and chemical abundance structure self-consistently
assuming steady-state chemistry.  Second we computed the chemical
abundances of gas and solid species for a 1~Myr old molecular cloud
with density of 5 $\times$ 10$^4$ cm$^{-3}$ and gas and dust
temperature of 15~K from diffuse cloud initial abundances, where the
elements are in neutral or ionized atomic form (see
Table~\ref{tab_initial_abundances}). Third we ran \ProDiMo in the
time-dependent chemical mode to simulate the chemical structure of a 1
Myr old disc using the results of the molecular cloud run as initial
chemical abundances and the disc properties computed in the first
stage. All other disc properties (UV field, density and temperature
structure) were fixed in this last stage.

The three-stage method mimics the incorporation of molecular cloud
materials and their subsequent evolution in the disc. It also makes it
possible to compare the chemical abundances obtained with the
time-dependent model and at steady-state. Our approach differs from
\citet{Visser2009A&A...495..881V} who solve the chemistry in a
Lagrangian frame and their disc evolves according to viscous
spreading. However their number of species and reactions are limited.

\begin{center}
\begin{table}
\caption{Stellar and disc parameters}\label{tab_DiscParameters}
		\begin{tabular}{lll}
\hline
Stellar mass & $M_*$ &  0.8~M$_\odot$ \\ 
Stellar luminosity &$L_*$  &  0.7~L$_\odot$ \\ 
Effective temperature & $T_{\mathrm {eff}}$ & 4400~K\\
Disc mass & $M_{\mathrm{d}}$ & 10$^{-3}$~M$_\odot$ \\
Disc inner radius & $R_{\mathrm {in}}$  & 0.1~AU\\
Disc outer radius & $R_{\mathrm {out}}$  & 300~AU\\
Vertical column density power law index & $\epsilon$ & 1\\
dust to gas mass ratio           &                     & 0.01\\
dust grain material mass density & $\rho_{\mathrm{dust}}$ & 2.5 g cm$^{-3}$ \\
minimum dust particle size       & $a_{\mathrm{min}}$ & 0.01 $\mu$m\\
maximum dust particle size       & $a_{\mathrm{max}}$ & 100 $\mu$m\\
dust size distribution power law & $p$               & 3.5\\
Cosmic ray flux                  & $CR$           & 1.7 $\times$ 10$^{-17}$ s$^{-1}$\\
ISM UV filed w.r.t. Draine field  & $\chi$          & 1.0\\
abundance of PAHs relative to ISM & $f_{\rm PAH}$      & 0.1\\
$\alpha$ viscosity parameter      & $\alpha$           & 0.0\\
\hline
\end{tabular}
\ \\
\end{table}
\end{center}
\begin{center}
\begin{table}
  \caption{Typical diffuse cloud abundances used as initial abundances for the molecular cloud chemical calculation. Species with
    ionization potential (IP) higher than 13.6 eV are neutral while species with IP below 13.6 eV are ionized. The other species have negligible initial abundance.} 
\label{tab_initial_abundances}
\begin{tabular}{ll}
\hline
Species & $\log$($n$(X)/$n_{\mathrm{H}}$) \\
\hline
H       & 0.0\\
D       & -5.0\\
He      & -1.125\\
C$^+$   & -3.886\\
O       & -3.538\\
N       & -4.67\\
S$^+$   & -5.721\\
Si$^+$  & -5.1\\
Mg$^+$  & -5.377\\
Fe$^+$  & -5.367\\
PAH     & -6.52\\
\hline
\end{tabular}
\ \\
\end{table}
\end{center}

\subsection{Model results}

\begin{figure}
\centering
\resizebox{\hsize}{!}
  {\includegraphics[angle=0]{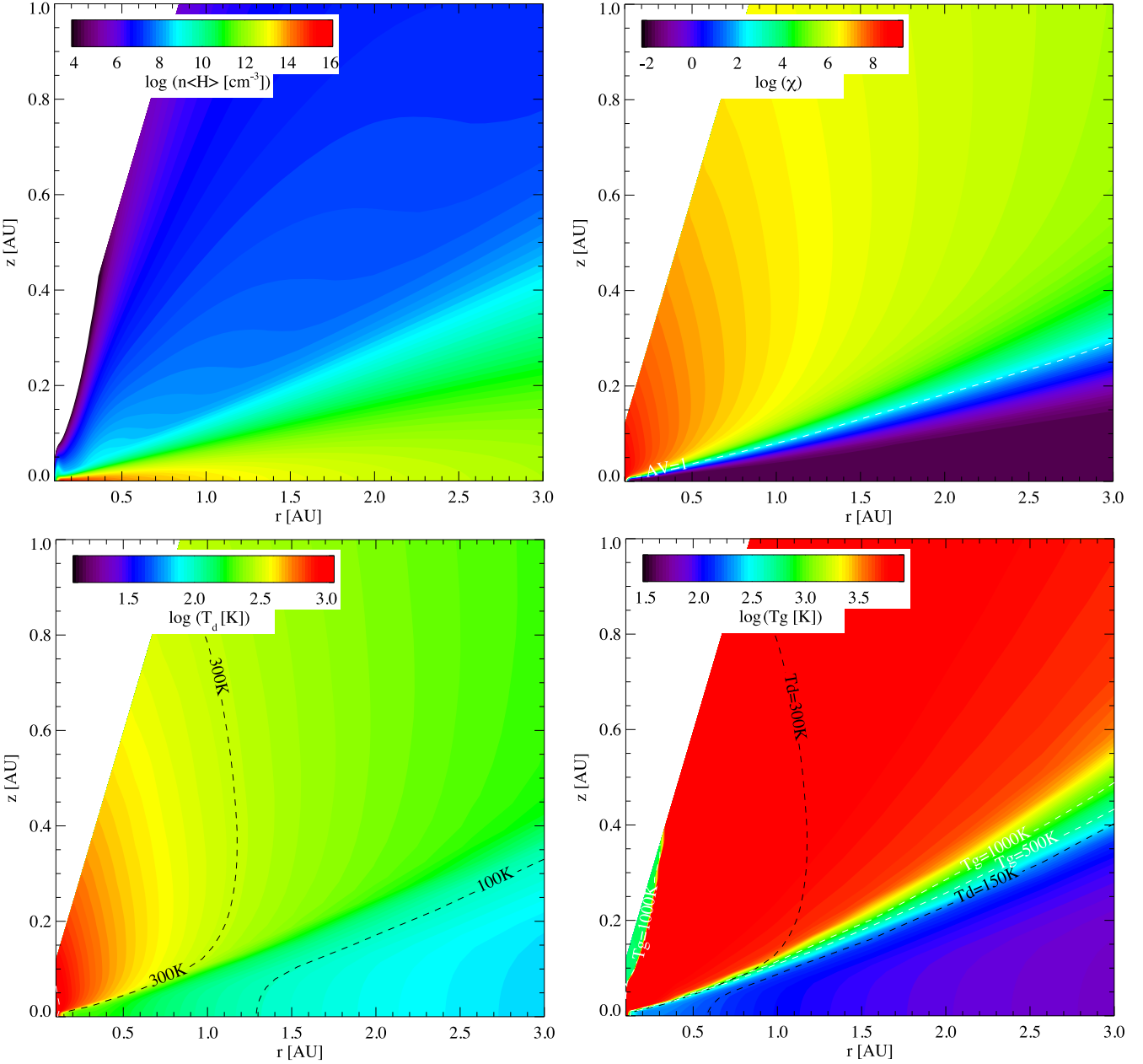}}
  \caption{The four panels show the gas density $n_{< {\mathrm H}>}$
    (upper left), strength of the UV field with respect to the Draine
    field $\chi$ (upper right), the dust temperature $T_{\mathrm{d}}$
    (lower left), and the gas temperature $T_{\mathrm{g}}$ with dust
    and gas temperature overlaid (lower right) for the inner 3~AU
    computed by {\sc ProDiMo}.  The structure is consistent with the
    chemical abundances.}
  \label{fig_nH_chi_Td_Tg}
\end{figure}

\begin{figure}
\centering
\resizebox{\hsize}{!}
  {\includegraphics[angle=0]{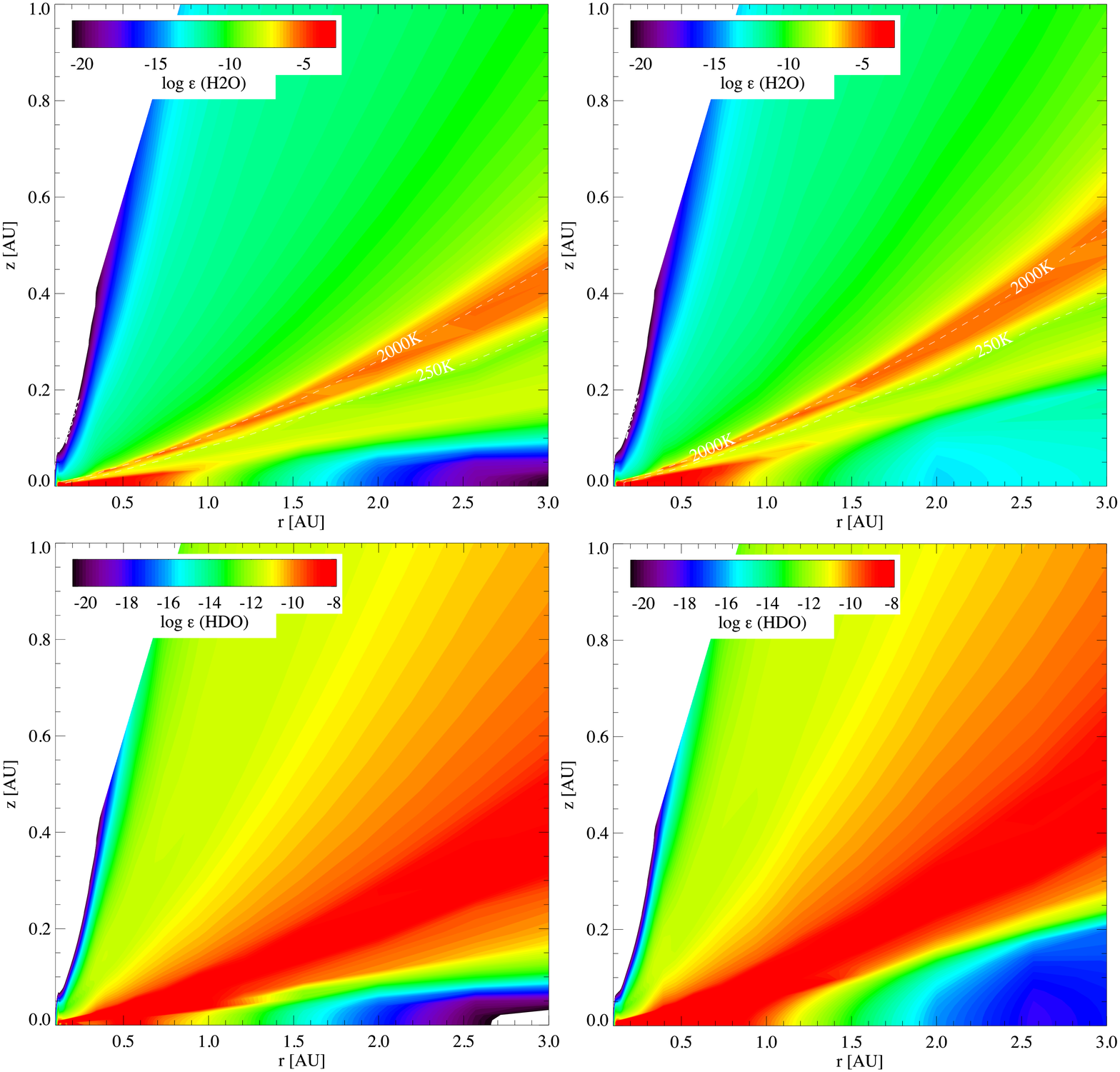}}
  \caption{Gas-phase H$_2$O and HDO abundance in the inner 3~AU
    computed by {\sc ProDiMo}. The left panels show the steady-state
    abundances while the right panels show the time-dependent
    abundances for a 1 Myr old disc.}
  \label{fig_steady_vs_time}
\end{figure}

Figure~\ref{fig_nH_chi_Td_Tg} show the UV field, density, and gas and
dust temperature structure in the inner 3~AU. The disc structure is
discussed in \citet{Woitke2009A&A...501..383W}. In this paper we focus
on the water and HDO abundances. Figure~\ref{fig_steady_vs_time} show
the water and HDO abundances at steady-state on the left and for a
disc of 1~Myr old on the right. Apart from the H$_2$O HDO in the
midplane beyond $\sim$1.5--2~AU, the water and HDO abundances are
similar for the steady-state and time-dependent chemistry models. The
differences stem from the fact that at steady-state even very slow
reactions will impact the chemistry.
In this case, the dust temperature is low enough ($T_{\mathrm
  d}<$~150~K) such that all water and HDO should be frozen-out. In the
time-dependent model, significant amount of water and HDO remains in
the gas-phase in the inner 1~AU or above the ice zone. 
The differences between the steady-state and time-dependent abundances
occur in low gas-phase water abundance regions and thus do not impede
our analysis of deuterium enrichment for gas-phase water either
outside the water freeze-out zone.

\begin{figure}
\centering
\resizebox{\hsize}{!}
  {\includegraphics[angle=0]{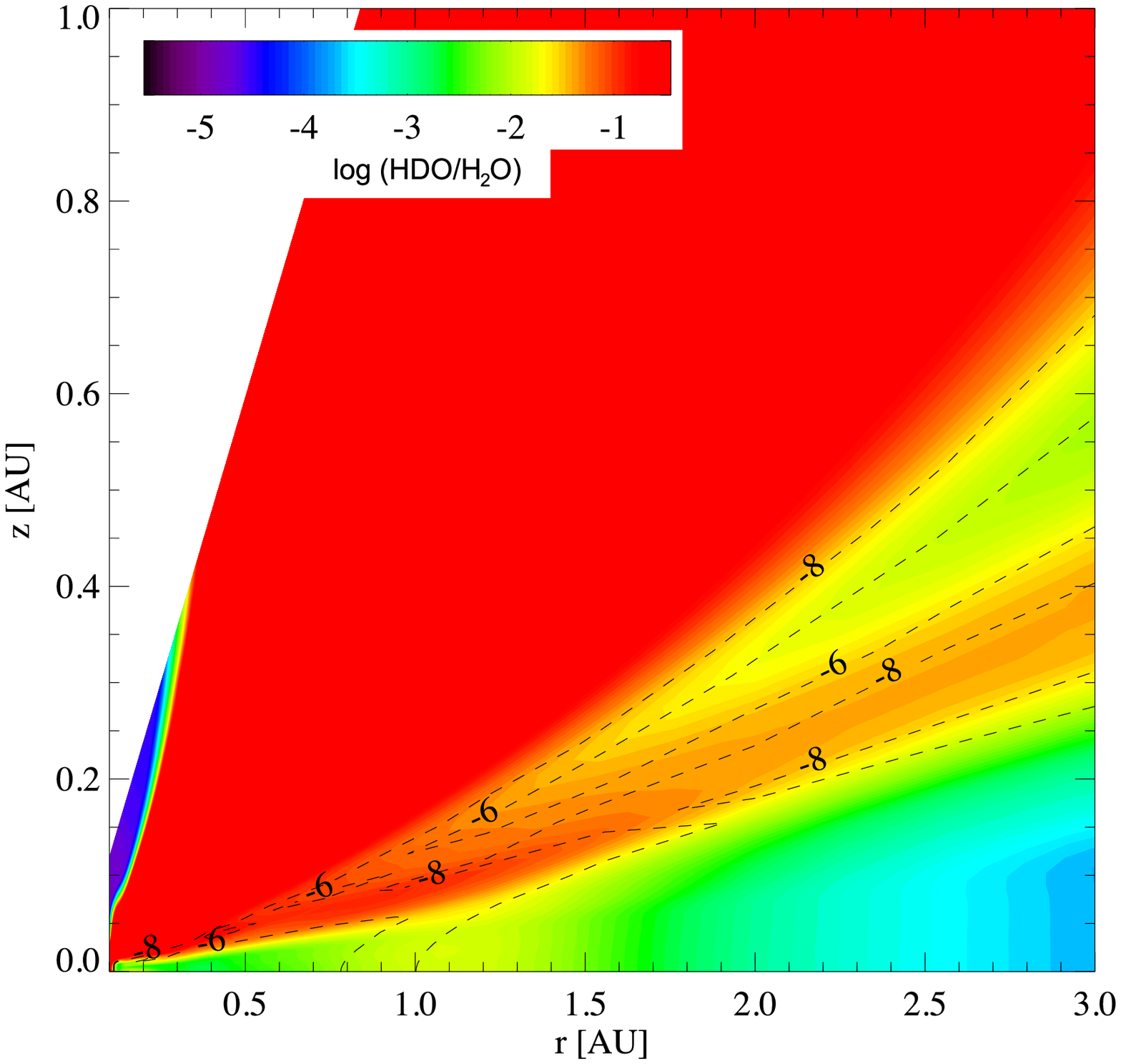}}
  \caption{[HDO]/[H$_2$O] in the 3~AU of a 10$^{-3}$ disc as computed
    by the photochemical code \ProDiMo. The contours indicate the regions where
    gas-phase water abundance is 10$^{-6}$ and 10$^{-8}$.}
  \label{fig_log_hdo_h2o}
\end{figure}

\begin{figure}
\centering
\resizebox{\hsize}{!}
  {\includegraphics[angle=0]{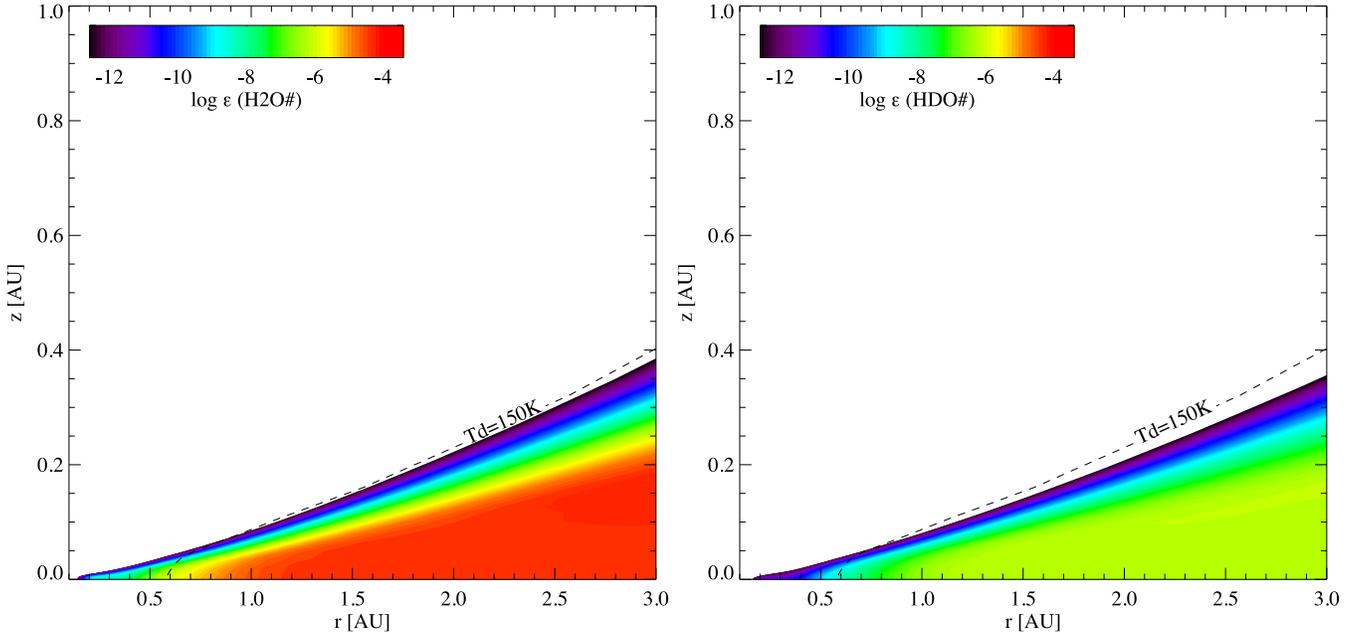}}
  \caption{H$_2$O and HDO ice abundance in the inner 3~AU. The
    dash-line indicate the region where the dust temperature is
    150~K. H$_2$O and HDO freeze at $\sim$150~K.}
  \label{fig_h2o_hdo_ice}
\end{figure}

\begin{figure}
\centering
\resizebox{\hsize}{!}
  {\includegraphics[angle=0]{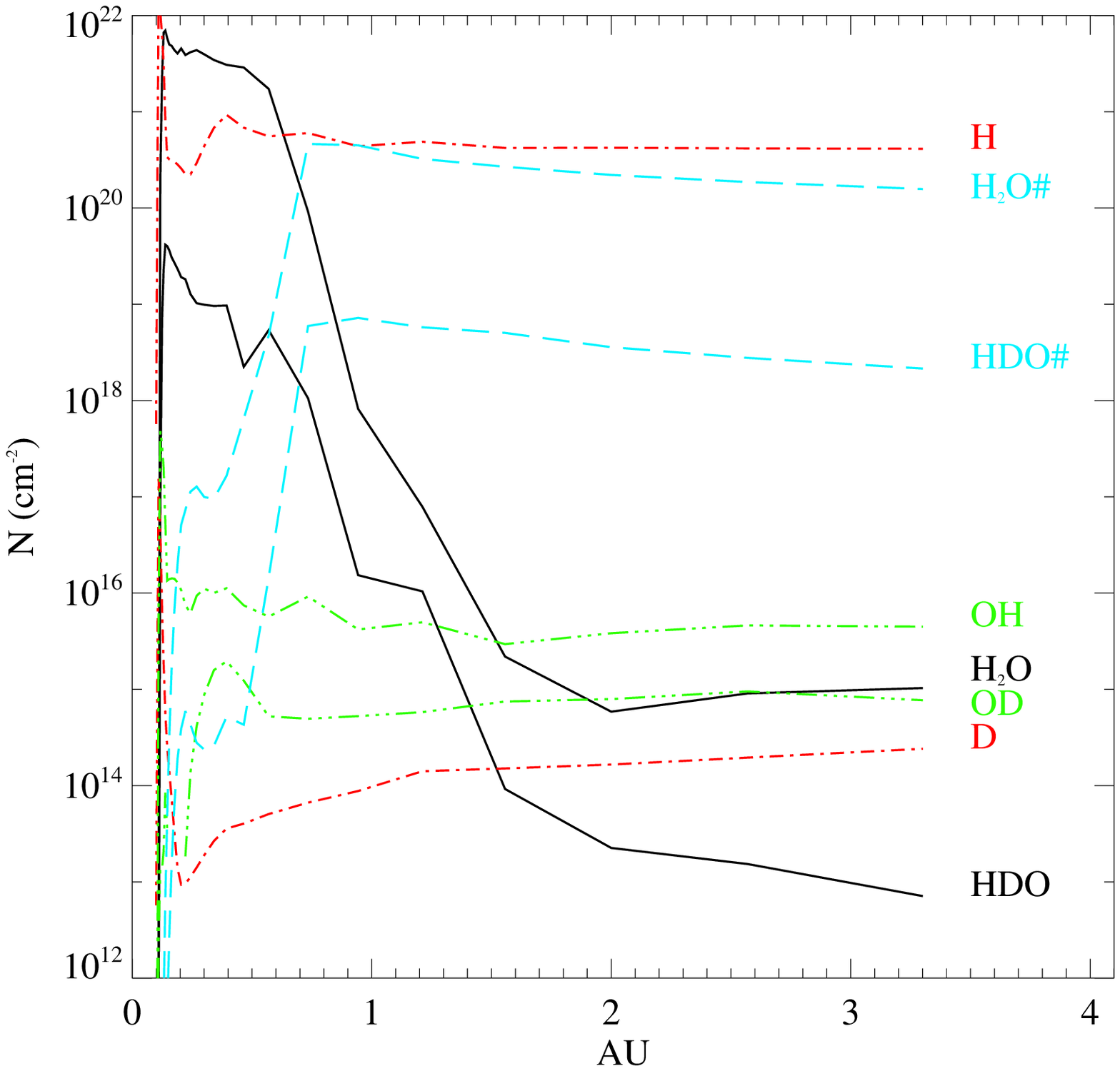}}
  \caption{Half disc vertical column density for various species in
    the inner 3~AU.}
  \label{fig_column_density}
\end{figure}

The [HDO]/[H$_2$O] ratio for the time-dependent model is plotted in
Figure~\ref{fig_log_hdo_h2o}.  The water abundance levels at 10$^{-6}$
and 10$^{-8}$ are overlaid. Water is abundant as soon as hydrogen is
in molecular form. In our model, water and HDO are frozen onto grain
surfaces in the midplane beyond $\sim$1.5~AU as shown in
Figure.~\ref{fig_h2o_hdo_ice}. The gas-phase [HDO]/[H$_2$O] ratio
decreases with radius. At 0.5--1.5~AU in the mid-plane, the ratio is a
few 10$^{-3}$--10$^{-2}$. 
Gas-phase water and HDO are located in three zones: the inner
midplane, the cold belt, and the hot layer
\citep{Woitke2009A&A...501L...5W}. The mass and average gas and dust
temperature are given in Table~\ref{tab_water_location}.
\begin{center}
\begin{table}
\caption{Mass and average gas and dust temperature for the three gas-phase water and HDO locations. The paper focuses
mostly on the inner midplane.}\label{tab_water_location}
		\begin{tabular}{llll}
\hline
Location         & \multicolumn{1}{c}{Mass}  & \multicolumn{1}{c}{$<T_{\mathrm{gas}}>$} & \multicolumn{1}{c}{$<T_{\mathrm{dust}}>$}\\
                 & \multicolumn{1}{c}{(M$_\odot$)} & \multicolumn{1}{c}{(K)} & \multicolumn{1}{c}{(K)}\\
\hline
& \multicolumn{3}{c}{H$_2$O}\\
\noalign{\smallskip} 
\cline{2-4}
\noalign{\smallskip} 
Inner midplane   &  2.6 $\times$ 10$^{-8}$ & 178 & 177\\
Cold belt        &  1.1 $\times$ 10$^{-9}$ & 16 & 16\\
Hot layer        &  2.1 $\times$ 10$^{-12}$ & 1147 & 120\\
\hline
& \multicolumn{3}{c}{HDO}\\
\noalign{\smallskip} 
\cline{2-4}
\noalign{\smallskip} 
Inner midplane   &  1.2 $\times$ 10$^{-10}$ & 140 & 140\\
Cold belt        &  3.8 $\times$ 10$^{-11}$ & 16 & 16\\
Hot layer        &  2.6 $\times$ 10$^{-14}$ & 474 & 102\\
\hline
\end{tabular}
\ \\
\end{table}
\end{center}
Most of the gas-phase water
(n(H$_2$O)/n$_{\mathrm{H}}$=10$^{-8}$--10$^{-5}$) is located in the
inner plane close to the star ($R<$1~AU). In the inner 30~AU a small
fraction is found above the mid-plane
\citep{Woitke2009A&A...501L...5W}, where the vertical extinction
$A_{\mathrm{V}}$ is between 1 and 10. Although of insignificant mass,
the molecules are hot and emit strongly (e.g.
\citealt{Carr2008Sci...319.1504C,Salyk2008ApJ...676L..49S}). In the
outer disc, gas-phase water is found in a cold belt, which is
sandwiched between $A_{\mathrm{V}}$ $\sim$1 and 5 where efficient
photodesorption maintains some water molecules in the gas phase (see
\citep{Woitke2009A&A...501L...5W}).  Since HDO and H$_2$O have similar
adsorption energy, the gas-phase abundance of both species are
co-located (see Fig.\ref{fig_steady_vs_time}). We plotted the vertical
column densities for H, D, OH, OD, H$_2$O, HDO, H$_2$O$\#$, and
HDO$\#$ in Fig.~\ref{fig_column_density}. This figure shows that the
column density ratios HDO/H$_2$O, HDO$\#$/H$_2$O$\#$, and OD/OH stay
relatively constant in the inner disc, with values much higher than
the elemental D/H ratio of 10$^{-5}$.

In the high water abundance region, the [HDO]/[H$_2$O] ratio is higher
than 10$^{-2}$. Using the mass of H$_2$O and HDO in the inner midplane
listed in Table~\ref{tab_water_location}, we derived an average
[HDO]/[H$_2$O] ratio of 4.6 $\times$ 10$^{-3}$, which is 30 times
higher than the Earth Mean Ocean Water value
($\simeq$~1.49$\times$10$^{-4}$) and close to cometary values.  The
actual [HDO]/[H$_2$O] values in our models depend on the disc
parameters. Future studies with focus on the effect of disc properties
(disc mass, radius, ...) on the [HDO]/[H$_2$O] ratios.

Overall, the [HDO]/[H$_2$O] ratios in the inner disc midplane by the
time-dependent code \ProDiMo are consistent with the analytical
results. Our results contradict the simple decrease in [H$_2$O]/[HDO]
with increasing temperature if thermochemical equilibrium ratios are assumed.

\section{Discussion and conclusion}\label{conlcusion}

A simple analytical analysis shows that high water fraction is
possible in a simple deuteration chemical network for gas hotter than
100 Kelvin if neutral-neutral reactions are included. The analytical
results are supported by the time-dependent photochemical model
results. In their study \citet{Willacy2009ApJ...703..479W} also found
that HDO can be abundant in the inner disc.

The water deuterium fractionation is determined by the ratio between
the rate of formation of OD via O + HD and the rate of formation of OH
via O + H$_2$. Lower zero-point energy for HD makes the first reaction
faster than the second one above 200~K. At low temperature
($T<$~100~K), fast deuteration of H$_3^+$, the main driver of cold
gas-phase chemistry ensures that main molecules are
deuterium-enriched. Another reason for higher HDO/H$_2$O is the
preferential branching into OD + H when HDO is photodissociated
\citep{Shafer1989,VanderWall1990,VanderWall1991}.  

From our work, it is also theoretically possible to have high water
deuterium fractionation from gas-phase photo-chemistry above
200~K. Deuterium enrichment can also occur at high temperature because
of the energy difference in activation barrier for deuterium exchange
and the back reactions.  Finally, water deuterium enrichment is not as
stringent a constraint as thought for the origin of water on
Earth. Deuterium-enriched water may have been synthesised at
$\sim$~1~AU and incorporated directly onto silicate dust grains
\citep{Stimpfl2006JCrGr.294...83S}. Water could have been incorporated
on Earth already at the stage of planetesimals accretion. However our
simple analytical steady-state chemical model and the 2D modelling
with {\sc ProDiMo} do not include effects of turbulent mixing in
protoplanetary discs, both radial and vertical.  

\section{Acknowledgments}
WFT is supported by a Scottish Universities Physics Alliance (SUPA)
fellowship in Astrobiology. We thank the referee for his/her comments.

\begin{center}
\begin{table}
\caption{Principal reactions}\label{tab_PrincipleReac}
		\begin{tabular}{lllllll}
	  \hline
 & Reaction & A & B & $E_{\mathrm a}$ &  Reference\\ 
& & (cm$^{3}$ s$^{-1}$) & & (K) & \\ 
\hline 
$k_1$ & O + \hh\ \ra\ OH + H & 3.14 $\times$ 10$^{-13}$ & 2.7 & 3150 & UMIST\\ 
$k_5$& OH + \hh\ \ra\ \water + H & 2.05 $\times$ 10$^{-12}$ & 1.52 & 1660 & UMIST \\ 
$k_{10}$ & \hh\ + D \ra\ HD + H & 7.50 $\times$ 10$^{-11}$ & -- & 3820 &  Zhang \& Millar 1989\\
$k_{11}$ & HD + H \ra\ \hh\ + D & 7.50 $\times$ 10$^{-11}$ & -- & 4240 &  Zhang \& Millar 1989\\ 
$k_{12}$ & O + HD \ra\ OD + H & 1.57 $\times$ 10$^{-12}$ & 1.7 & 4639 &  Joseph, Truhlar \& Garret 1988 \\ 
$k_{16}$ & OH + D \ra\ OD + H & 9.07 $\times$ 10$^{-11}$ & -0.63 & -- &  Yung et al. 1988\\ 
$k_{20}$ & OH + HD \ra\ HDO + H & 0.60 $\times$ 10$^{-13}$ & 1.9 & 1258 &  Talukdar et al.\ 1996\\
$k_{22}$ & OD + \hh\ \ra\ HDO + H & 1.55 $\times$ 10$^{-12}$ & 1.6 & 1663 & Talukdar et al.\ 1996\\ 		
\hline
 & Reaction      &  q &  $\gamma$ & & Reference\\
 &              &          & (s$^{-1}$) & & & \\ 
\hline 
$J_2$ & HD + \UV\ \ra\ H + D & 2.6 $\times$ 10$^{-11}$& 2.5 & -- & Le Petit et al.\ 2002\\
$J_4$ & OH + \UV\ \ra\ O + H & 3.5 $\times$ 10$^{-10}$ & 1.7 & -- & van Dishoeck 1988 \\ 
\hline
\end{tabular}
\ \\
Ref. UMIST: \citet{Wooddall2007A&A...466.1197W}
\end{table}
\end{center}

\begin{center}
\begin{table}
\caption[]{Neutral-neutral and radical-neutral reactions}\label{tab_Reactions1}
\begin{tabular}{lllllll}
  \hline 
 & Reaction & A & B & $E_{\mathrm a}$  &  Reference\\ 
& & (cm$^{3}$ s$^{-1}$) & (K) & (K) & \\ 
\hline 
$R_1$ & H + H + M \ra\ \hh & -- & -- & -- & see text \\ 
$R_{2}$ & H + D + M \ra\ HD & -- & -- & -- & see text \\ 
\hline
$k_1$ & O + \hh\ \ra\ OH + H & 3.14 $\times$ 10$^{-13}$ &2.70 &3150 &  UMIST\\ 
$k_2$& O$_2$ + H \ra\ OH + H & 2.61 $\times$ 10$^{-10}$ & -- & 8156 &  UMIST\\ 
$k_3$& O$_2$ + \hh\ \ra\ OH + OH & 3.16 $\times$ 10$^{-10}$ & -- & 21890 &  UMIST\\ 
$k_4$& OH + H \ra\ O + \hh &  7.00 $\times$ 10$^{-14}$ & 2.80 & 1950 & UMIST \\ 
$k_5$& OH + \hh\ \ra\ \water + H & 2.05 $\times$ 10$^{-12}$ & 1.52 & 1660 & UMIST\\ 
$k_6$& OH + O \ra\ O$_2$ + H & 1.77 $\times$ 10$^{-11}$ & -- & -178 &  UMIST\\ 
$k_7$& OH + OH \ra\ O + \water & 3.87 $\times$ 10$^{-13}$ & 1.69 & -469 &  UMIST\\ 
$k_8$& \water\ + H \ra\ OH + \hh & 1.59 $\times$ 10$^{-11}$ & 1.20 & 9610  & UMIST\\ 
$k_9$& \water + O \ra\ OH + OH & 1.85 $\times$ 10$^{-11}$ & 0.95 & 8571 &  UMIST\\ 
$k_{10}$ & \hh\ + D \ra\ HD + H & 7.50 $\times$ 10$^{-11}$ & -- & 3820 &  Zhang \& Millar 1989\\ 
$k_{11}$ & HD + H \ra\ \hh\ + D & 7.50 $\times$ 10$^{-11}$ & -- & 4240 &  Zhang \& Millar 1989\\ 
$k_{12}$ & O + HD \ra\ OD + H & 1.57 $\times$ 10$^{-12}$ & 1.7 & 4639 &  Joseph, Truhlar \& Garret 1988 \\ 
$k_{13}$ & OD + H \ra\ O + HD & -- & -- & -- & $k_{13}=k_4$ is assumed\\ 
$k_{14}$ & O + HD \ra\ OH + D & 9.01 $\times$ 10$^{-13}$ & 1.9 & 3730 &  Joseph, Truhlar \& Garret 1988 \\ 
$k_{15}$ & OH + D \ra\ O + HD & -- & -- & -- & $k_{15}=k_4$ is assumed\\ 
$k_{16}$ & OH + D \ra\ OD + H & 9.07 $\times$ 10$^{-11}$ & -0.63 & -- &  Yung et al. 1988\\ 
$k_{17}$ & OD + H \ra\ OH + D & 1.26 $\times$ 10$^{-10}$ & -0.63 & 717 &  Yung et al. 1988\\ 
$k_{18}$ & OH + HD \ra\ \water\ + D & 2.12 $\times$ 10$^{-13}$ & 2.7 & 1258 &  Talukdar et al.\ 1996\\ 
$k_{19}$ & \water\ + D \ra\ OH + HD &-- &-- &-- & $k_{19}=k_8$ is assumed\\ 
$k_{20}$ & OH + HD \ra\ HDO + H & 0.60 $\times$ 10$^{-13}$ & 1.9 & 1258 &  Talukdar et al.\ 1996\\ 
$k_{21}$ & HDO + H \ra\ OH + HD & -- & -- & -- & $k_{21}=0.5 \times k_8$ is assumed\\ 
$k_{22}$ & OD + \hh\ \ra\ HDO + H & 1.55 $\times$ 10$^{-12}$ & 1.6 & 1663 &  Talukdar et al.\ 1996\\ 
$k_{23}$ & HDO + H \ra\ OD + \hh\ & -- & -- & -- & $k_{23}=0.5 \times k_8$ is assumed\\ 
$k_{24}$ & OD + O \ra\ O$_2$ + D & -- & -- & -- & $k_{24}=k_{6}$ is assumed \\ 
$k_{25}$ & O$_2$ + D \ra\ OD + O & -- & -- & -- & $k_{25}=k_{2}$ is assumed\\ 
$k_{26}$ & OD + OH \ra\ O + HDO & -- & -- & -- & $k_{26}=k_{7}$ is assumed \\ 
$k_{27}$ & HDO + O \ra\ OD + OH & -- & -- & -- & $k_{27}=k_{9}$ is assumed\\ 
\hline
$k_{28}$  & OH + CO \ra\ CO$_2$ + H & 1.17 $\times$ 10$^{-13}$ & 0.95 & -74.0 & UMIST\\ 
$k_{29}$ & OH + C \ra\ CO + H  & 1.10 $\times$ 10$^{-10}$ & 0.5 & 0.0 & UMIST\\ 
$k_{30}$ & CO + H \ra\ OH + C & 1.10 $\times$ 10$^{-10}$ &0.5 &77700.0 & UMIST\\ 
\hline
$k_{31}$ & OD + CO \ra\ CO$_2$ + D & -- & -- & -- & $k_{31}=k_{28}$ is assumed \\ 
$k_{32}$ & OD + C \ra\ CO + D      & -- & -- & -- & $k_{32}=k_{29}$ is assumed \\ 
$k_{33}$ & CO + D \ra\ OD + C      & -- & -- & -- & $k_{33}=k_{30}$ is assumed \\ 
\hline
\end{tabular}
\ \\
Ref. UMIST: \citet{Wooddall2007A&A...466.1197W}
\end{table}
\end{center}

\begin{center}
\begin{table} 
\caption[]{Photodissociation and photoionisation reactions}	\label{tab_Reactions2}
\begin{tabular}{llllllll}
  \hline 
& Reaction & q &  $\gamma$ & Reference\\
&          & (s$^{-1}$) & & \\ 
\hline 
$J_1$ & \hh\ + \UV\ \ra\ H + H & 3.4 $\times$ 10$^{-11}$ & 2.5  & self-shielding factor, see text\\ 
$J_2$ & HD + \UV\ \ra\ H + D & 2.6 $\times$ 10$^{-11}$ & 2.5 & LePetit et al.\ 2002\\
$J_3$ & CO + \UV\ \ra\ C + O & 2.0 $\times$ 10$^{-10}$ & 2.5 & UMIST\\
$J_4$ & OH + \UV\ \ra\ O + H & 3.5 $\times$ 10$^{-10}$ & 1.7 & van Dishoeck 1988\\ 
$J_5$ & OD + \UV\ \ra\ O + D & 4.0 $\times$ 10$^{-10}$ & 1.7 & Croswell \& Dalgarno 1985\\ 
$J_6$ & \water\ + \UV\ \ra\ H + OH & 5.9 $\times$ 10$^{-10}$ & 1.7 & UMIST \\ 
$J_{7a}$ & HDO + \UV\ \ra\ OH + D & -- & -- & $J_{7a}=0.25 J_6$, see text\\ 
$J_{7b}$ & HDO + \UV\ \ra\ OD + H & -- & -- & $J_{7b}=0.75 J_6$, see text\\ 
\hline
$J_9$ & C + \UV\ \ra\ C$^+$ + e & & & UMIST \\ 
\hline
\end{tabular}
\ \\
Ref. UMIST: \citet{Wooddall2007A&A...466.1197W}
\end{table}
\end{center}
\begin{center}
\begin{table}
\caption[]{Ion-neutral reactions}\label{tab_Reactions3}
\begin{tabular}{llll}
  \hline 
 & Reaction & $k$ &  Reference\\ 
& & (cm$^{3}$ s$^{-1}$) & \\ 
\hline
$k_{34}$ & H$^+$ + D \ra\ D$^+$ + H & 1.0 $\times$ 10$^{-9}\ e^{-41/T}$ &  \citet{Watson1976RvMP...48..513W}, UMIST\\	
$k_{35}$ & He$^+$ + H$_2$ \ra\ H + H$^+$ + He & 1.1 $\times$ 10$^{-13}(T/300)^{-0.24}$& UMIST\\
$k_{36}$ & HD + H$^+$ \ra\ D$^+$ + H$_2$ & 1.0 $\times$ 10$^{-9}\ e^{-464/T}$ & UMIST \\
$k_{37}$ & H$_2$ + D$^+$ \ra\ H$^+$ +HD  & 2.1 $\times$ 10$^{-9}$ & UMIST\\
$k_{38}$ & H + D$^+$ \ra\ H$^+$ + D & 1.0 $\times$ 10$^{-9}$ & \citet{Watson1976RvMP...48..513W}, UMIST\\ 
$k_{39}$ & OH + C$^+$ \ra\ CO$^+$ + H & 7.7 $\times$10$^{-10}$& \citet{Prasas1980ApJS...43....1P}, UMIST\\ 
$k_{40}$ & OD + C$^+$ \ra\ CO$^+$ + D & -- & $k_{40}=k_{39}$ is assmued\\ 
$k_{41}$ & He$^+$ + H$_2$O \ra\ OH + He + H$^+$ & 6.0 $\times$ 10$^{-11}$ & UMIST\\
$k_{42}$ & He$^+$ + HDO \ra\ OD + He + H$^+$ & -- & $k_{42}=0.75k_{41}$ is assumed\\
$k_{43}$ & He$^+$ + HDO \ra\ OH + He + D$^+$ & -- & $k_{43}=0.25k_{41}$ is assumed\\
$k_{44}$ & He$^+$ + OH \ra\ O$^+$ + He + H & 1.1 $\times$ 10$^{-9}$ & UMIST\\
$k_{45}$ & He$^+$ + OD \ra\ O$^+$ + He + D & -- & $k_{45}=k_{44}$ is assmued\\
\hline
\end{tabular}
\ \\
Ref. UMIST: \citet{Wooddall2007A&A...466.1197W}
\end{table}
\end{center}
\begin{center}
\begin{table}
\caption[]{Cosmic-ray induced ionisation and recombination reactions ($\zeta = 5 \times 10^{-17}$ s$^{-1}$).}\label{tab_Reactions4}
\begin{tabular}{llll}
  \hline 
 & Reaction & $k$ &  Reference\\ 
& & (cm$^{3}$ s$^{-1}$) & \\ 
\hline
$k_{\zeta,1}$ & H + cr \ra\ H$^+$ + e$^-$ & 0.46 $\times$ $\zeta$ & UMIST \\
$k_{\zeta,2}$ & D + cr \ra\ D$^+$ + e$^-$ & 0.46 $\times$ $\zeta$ & UMIST \\
$k_{\zeta,3}$ & He + cr \ra\ He$^+$ + e$^-$ & 0.5 $\times$ $\zeta$ & UMIST \\
$k_{\zeta,4}$ & H$_2$ + cr \ra\ H$^+$ + H + e$^-$ & 0.04 $\times$ $\zeta$ & UMIST \\
$k_{\zeta,5}$ & HD + cr \ra\ H$^+$ + D + e$^-$ & -- & $k_{\zeta,5}=0.5 k_{\zeta,4}$ is assumed\\
$k_{\zeta,6}$ & HD + cr \ra\ D$^+$ + H + e$^-$ & -- & $k_{\zeta,6}=0.5 k_{\zeta,4}$ is assumed\\
$k_{e^-,1}$ & H$^+$ + e$^-$ \ra\ H + \UV\ & $3.5 \times 10^{-12}(T/300)^{-0.70}$& \citet{Prasas1980ApJS...43....1P}\\
$k_{e^-,2}$ & D$^+$ + e$^-$ \ra\ D + \UV\ & -- & $k_{e^-,2}= k_{e^-,1}$ is assumed\\
$k_{e^-,3}$ & He$^+$ + e$^-$ \ra\ He + \UV\ & $4.5 \times 10^{-12}(T/300)^{-0.67}$& \citet{Prasas1980ApJS...43....1P}\\
$k_{cp,1}$  & OH + CRPhot \ra\ O + H & 1.3 $\times$ 10$^{-17} (509/(1-w))$& UMIST \\
$k_{cp,2}$  & OD + CRPhot \ra\ O + D & 1.3 $\times$ 10$^{-17} (509/(1-w))$& $k_{cp,2}=k_{cp,1}$ is assumed \\
$k_{cp,3}$ & H$_2$O + CRPhot \ra\ OH + H &  1.3 $\times$ 10$^{-17} (971/(1-w))$& UMIST \\
$k_{cp,4}$ & HDO + CRPhot \ra\ OD + H &  $0.75 \times 1.3 \times 10^{-17} (971/(1-w))$ & UMIST \\
$k_{cp,5}$ & HDO + CRPhot \ra\ OH + D &  $0.25 \times 1.3 \times 10^{-17} (971/(1-w))$ & UMIST \\
\hline
\end{tabular}
\ \\
Ref. UMIST: \citet{Wooddall2007A&A...466.1197W}
\end{table}
\end{center}


\bibliographystyle{mn2e}
\bibliography{hdo_h2o}

\begin{thebibliography}{}

\bibitem[\protect\citeauthoryear{{Bergin}, {Neufeld} \& {Melnick}}{{Bergin}
  et~al.}{1998}]{Bergin1998ApJ...499..777B}
{Bergin} E.~A.,  {Neufeld} D.~A.,    {Melnick} G.~J.,  1998, \apj, 499, 777

\bibitem[\protect\citeauthoryear{{Brott} \& {Hauschildt}}{{Brott} \&
  {Hauschildt}}{2005}]{Brott2005ESASP.576..565B}
{Brott} I.,  {Hauschildt} P.~H.,  2005, in {C.~Turon, K.~S.~O'Flaherty, \&
  M.~A.~C.~Perryman} ed., The Three-Dimensional Universe with Gaia Vol.~576 of
  ESA Special Publication, {A PHOENIX Model Atmosphere Grid for Gaia}.
pp 565--+

\bibitem[\protect\citeauthoryear{{Brown} \& {Millar}}{{Brown} \&
  {Millar}}{1989}]{Brown1989MNRAS.237..661B}
{Brown} P.~D.,  {Millar} T.~J.,  1989, \mnras, 237, 661

\bibitem[\protect\citeauthoryear{{Carr} \& {Najita}}{{Carr} \&
  {Najita}}{2008}]{Carr2008Sci...319.1504C}
{Carr} J.~S.,  {Najita} J.~R.,  2008, Science, 319, 1504

\bibitem[\protect\citeauthoryear{{Cazaux} \& {Tielens}}{{Cazaux} \&
  {Tielens}}{2002}]{Cazaux2002ApJ...575L..29C}
{Cazaux} S.,  {Tielens} A.~G.~G.~M.,  2002, \apjl, 575, L29

\bibitem[\protect\citeauthoryear{{Ceccarelli}, {Dominik}, {Caux}, {Lefloch} \&
  {Caselli}}{{Ceccarelli} et~al.}{2005}]{Ceccarelli2005ApJ...631L..81C}
{Ceccarelli} C.,  {Dominik} C.,  {Caux} E.,  {Lefloch} B.,    {Caselli} P.,
  2005, \apjl, 631, L81

\bibitem[\protect\citeauthoryear{{Charnley}, {Tielens} \& {Rodgers}}{{Charnley}
  et~al.}{1997}]{Charnley1997ApJ...482L.203C}
{Charnley} S.~B.,  {Tielens} A.~G.~G.~M.,    {Rodgers} S.~D.,  1997, \apjl,
  482, L203+

\bibitem[\protect\citeauthoryear{{Croswell} \& {Dalgarno}}{{Croswell} \&
  {Dalgarno}}{1985}]{Croswell1985ApJ...289..618C}
{Croswell} K.,  {Dalgarno} A.,  1985, \apj, 289, 618

\bibitem[\protect\citeauthoryear{{Dartois}, {Thi}, {Geballe}, {Deboffle},
  {d'Hendecourt} \& {van Dishoeck}}{{Dartois}
  et~al.}{2003}]{Dartois2003A&A...399.1009D}
{Dartois} E.,  {Thi} W.-F.,  {Geballe} T.~R.,  {Deboffle} D.,  {d'Hendecourt}
  L.,    {van Dishoeck} E.,  2003, \aap, 399, 1009

\bibitem[\protect\citeauthoryear{{Dorren} \& {Guinan}}{{Dorren} \&
  {Guinan}}{1994}]{Dorren1994ApJ...428..805D}
{Dorren} J.~D.,  {Guinan} E.~F.,  1994, \apj, 428, 805

\bibitem[\protect\citeauthoryear{{Draine} \& {Bertoldi}}{{Draine} \&
  {Bertoldi}}{1996}]{Draine1996ApJ...468..269D}
{Draine} B.~T.,  {Bertoldi} F.,  1996, \apj, 468, 269

\bibitem[\protect\citeauthoryear{{Drake}}{{Drake}}{2005}]{Drake2005M&PS...40..%
519D}
{Drake} M.~J.,  2005, Meteoritics and Planetary Science, 40, 519

\bibitem[\protect\citeauthoryear{{Genda} \& {Ikoma}}{{Genda} \&
  {Ikoma}}{2008}]{Genda2008Icar..194...42G}
{Genda} H.,  {Ikoma} M.,  2008, Icarus, 194, 42

\bibitem[\protect\citeauthoryear{{Gensheimer}, {Mauersberger} \&
  {Wilson}}{{Gensheimer} et~al.}{1996}]{Gensheimer1996A&A...314..281G}
{Gensheimer} P.~D.,  {Mauersberger} R.,    {Wilson} T.~L.,  1996, \aap, 314,
  281

\bibitem[\protect\citeauthoryear{{Glassgold}, {Meijerink} \&
  {Najita}}{{Glassgold} et~al.}{2009}]{Glassgold2009ApJ...701..142G}
{Glassgold} A.~E.,  {Meijerink} R.,    {Najita} J.~R.,  2009, \apj, 701, 142

\bibitem[\protect\citeauthoryear{{Gomes}, {Levison}, {Tsiganis} \&
  {Morbidelli}}{{Gomes} et~al.}{2005}]{Gomes2005Natur.435..466G}
{Gomes} R.,  {Levison} H.~F.,  {Tsiganis} K.,    {Morbidelli} A.,  2005, \nat,
  435, 466

\bibitem[\protect\citeauthoryear{{Guilloteau}, {Pi{\'e}tu}, {Dutrey} \&
  {Gu{\'e}lin}}{{Guilloteau} et~al.}{2006}]{Guilloteau2006A&A...448L...5G}
{Guilloteau} S.,  {Pi{\'e}tu} V.,  {Dutrey} A.,    {Gu{\'e}lin} M.,  2006,
  \aap, 448, L5

\bibitem[\protect\citeauthoryear{{Hopkins}, {Harrison} \& {Manning}}{{Hopkins}
  et~al.}{2008}]{Hopkins2008Nature}
{Hopkins} M.,  {Harrison} T.~M.,    {Manning} C.~E.,  2008, Nature, 456, 493

\bibitem[\protect\citeauthoryear{{Kamp}, {Tilling}, {Woitke}, {Thi} \&
  {Hogerheijde}}{{Kamp} et~al.}{2009}]{Kamp2009}
{Kamp} I.,  {Tilling} I.,  {Woitke} P.,  {Thi} W.,    {Hogerheijde} M.,  2009,
  ArXiv e-prints

\bibitem[\protect\citeauthoryear{{Le Petit}, {Roueff} \& {Le Bourlot}}{{Le
  Petit} et~al.}{2002}]{LePetit2002A&A...390..369L}
{Le Petit} F.,  {Roueff} E.,    {Le Bourlot} J.,  2002, \aap, 390, 369

\bibitem[\protect\citeauthoryear{{Linsky}}{{Linsky}}{2003}]{Linsky2003SSRv..10%
6...49L}
{Linsky} J.~L.,  2003, Space Science Reviews, 106, 49

\bibitem[\protect\citeauthoryear{{Lyons} \& {Young}}{{Lyons} \&
  {Young}}{2005}]{Lyons2005Natur.435..317L}
{Lyons} J.~R.,  {Young} E.~D.,  2005, \nat, 435, 317

\bibitem[\protect\citeauthoryear{{Morbidelli}, {Chambers}, {Lunine}, {Petit},
  {Robert}, {Valsecchi} \& {Cyr}}{{Morbidelli}
  et~al.}{2000}]{Morbidelli2000M&PS...35.1309M}
{Morbidelli} A.,  {Chambers} J.,  {Lunine} J.~I.,  {Petit} J.~M.,  {Robert} F.,
   {Valsecchi} G.~B.,    {Cyr} K.~E.,  2000, Meteoritics and Planetary Science,
  35, 1309

\bibitem[\protect\citeauthoryear{{Nuth}}{{Nuth}}{2008}]{Nuth2008EM&P..102..435%
N}
{Nuth} J.~A.,  2008, Earth Moon and Planets, 102, 435

\bibitem[\protect\citeauthoryear{{Parise}, {Caux}, {Castets}, {Ceccarelli},
  {Loinard}, {Tielens}, {Bacmann}, {Cazaux}, {Comito}, {Helmich}, {Kahane},
  {Schilke}, {van Dishoeck}, {Wakelam} \& {Walters}}{{Parise}
  et~al.}{2005}]{Parise2005A&A...431..547P}
{Parise} B.,  {Caux} E.,  {Castets} A.,  {Ceccarelli} C.,  {Loinard} L.,
  {Tielens} A.~G.~G.~M.,  {Bacmann} A.,  {Cazaux} S.,  {Comito} C.,  {Helmich}
  F.,  {Kahane} C.,  {Schilke} P.,  {van Dishoeck} E.,  {Wakelam} V.,
  {Walters} A.,  2005, \aap, 431, 547

\bibitem[\protect\citeauthoryear{{Parise}, {Simon}, {Caux}, {Dartois},
  {Ceccarelli}, {Rayner} \& {Tielens}}{{Parise}
  et~al.}{2003}]{Parise2003A&A...410..897P}
{Parise} B.,  {Simon} T.,  {Caux} E.,  {Dartois} E.,  {Ceccarelli} C.,
  {Rayner} J.,    {Tielens} A.~G.~G.~M.,  2003, \aap, 410, 897

\bibitem[\protect\citeauthoryear{{Prasad} \& {Huntress} Jr.}{{Prasad} \&
  {Huntress}}{1980}]{Prasas1980ApJS...43....1P}
{Prasad} S.~S.,  {Huntress} Jr. W.~T.,  1980, \apjs, 43, 1

\bibitem[\protect\citeauthoryear{{Raymond}, {Quinn} \& {Lunine}}{{Raymond}
  et~al.}{2004}]{Raymond2004Icar..168....1R}
{Raymond} S.~N.,  {Quinn} T.,    {Lunine} J.~I.,  2004, Icarus, 168, 1

\bibitem[\protect\citeauthoryear{{Raymond}, {Quinn} \& {Lunine}}{{Raymond}
  et~al.}{2005}]{Raymond2005ApJ...632..670R}
{Raymond} S.~N.,  {Quinn} T.,    {Lunine} J.~I.,  2005, \apj, 632, 670

\bibitem[\protect\citeauthoryear{{Richet}, {Bottinga} \& {Janoy}}{{Richet}
  et~al.}{1977}]{Richet1977AREPS...5...65R}
{Richet} P.,  {Bottinga} Y.,    {Janoy} M.,  1977, Annual Review of Earth and
  Planetary Sciences, 5, 65

\bibitem[\protect\citeauthoryear{{Righter}, {Drake} \& {Scott}}{{Righter}
  et~al.}{2006}]{Righter2006mess.book..803R}
{Righter} K.,  {Drake} M.~J.,    {Scott} E.~R.~D.,  2006, {Compositional
  Relationships Between Meteorites and Terrestrial Planets}.
Meteorites and the Early Solar System II, pp 803--828

\bibitem[\protect\citeauthoryear{{Robert}, {Gautier} \& {Dubrulle}}{{Robert}
  et~al.}{2000}]{Robert2000SSRv...92..201R}
{Robert} F.,  {Gautier} D.,    {Dubrulle} B.,  2000, Space Science Reviews, 92,
  201

\bibitem[\protect\citeauthoryear{{Roberts}, {Herbst} \& {Millar}}{{Roberts}
  et~al.}{2004}]{Roberts2004A&A...424..905R}
{Roberts} H.,  {Herbst} E.,    {Millar} T.~J.,  2004, \aap, 424, 905

\bibitem[\protect\citeauthoryear{{Roberts} \& {Millar}}{{Roberts} \&
  {Millar}}{2000}]{Roberts2000A&A...361..388R}
{Roberts} H.,  {Millar} T.~J.,  2000, \aap, 361, 388

\bibitem[\protect\citeauthoryear{{Salyk}, {Pontoppidan}, {Blake}, {Lahuis},
  {van Dishoeck} \& {Evans} II}{{Salyk}
  et~al.}{2008}]{Salyk2008ApJ...676L..49S}
{Salyk} C.,  {Pontoppidan} K.~M.,  {Blake} G.~A.,  {Lahuis} F.,  {van Dishoeck}
  E.~F.,    {Evans} II N.~J.,  2008, \apjl, 676, L49

\bibitem[\protect\citeauthoryear{{Shafer}, {Satyapal} \& {Bersohn}}{{Shafer}
  et~al.}{1989}]{Shafer1989}
{Shafer} N.,  {Satyapal} S.,    {Bersohn} R.,  1989, J. Chem Phys., 90, 6807

\bibitem[\protect\citeauthoryear{{Stimpfl}, {Walker}, {Drake}, {de Leeuw} \&
  {Deymier}}{{Stimpfl} et~al.}{2006}]{Stimpfl2006JCrGr.294...83S}
{Stimpfl} M.,  {Walker} A.~M.,  {Drake} M.~J.,  {de Leeuw} N.~H.,    {Deymier}
  P.,  2006, Journal of Crystal Growth, 294, 83

\bibitem[\protect\citeauthoryear{{Thi} \& {Bik}}{{Thi} \&
  {Bik}}{2005}]{Thi2005A&A...438..557T}
{Thi} W.-F.,  {Bik} A.,  2005, \aap, 438, 557

\bibitem[\protect\citeauthoryear{{Tielens}}{{Tielens}}{2005}]{Tielens2005pcim.%
book.....T}
{Tielens} A.~G.~G.~M.,  2005, {The Physics and Chemistry of the Interstellar
  Medium}.
The Physics and Chemistry of the Interstellar Medium, by A.~G.~G.~M.~Tielens,
  pp.~.~ISBN 0521826349.~Cambridge, UK: Cambridge University Press, 2005.

\bibitem[\protect\citeauthoryear{{van Dishoeck}}{{van
  Dishoeck}}{1988}]{vanDishoeck1988rcia.conf...49V}
{van Dishoeck} E.~F.,  1988, in {Millar} T.~J.,  {Williams} D.~A.,  eds, Rate
  Coefficients in Astrochemistry. Proceedings of a Conference held in UMIST,
  Manchester, United Kingdom, September 21-24, 1987. Editors, T.J. Millar, D.A.
  Williams; Publisher, Kluwer Academic Publishers, Dordrecht, Boston, 1988.
  ISBN \# 90-277-2752-X. LC \# QB450 .R38 1988. P. 49, 1988 {Photodissociation
  and Photoionization Processes}.
pp 49--+

\bibitem[\protect\citeauthoryear{{van Dishoeck} \& {Dalgarno}}{{van Dishoeck}
  \& {Dalgarno}}{1984}]{vanDishoeck1984ApJ...277..576V}
{van Dishoeck} E.~F.,  {Dalgarno} A.,  1984, \apj, 277, 576

\bibitem[\protect\citeauthoryear{{van Dishoeck}, {Jonkheid} \& {van
  Hemert}}{{van Dishoeck} et~al.}{2008}]{vanDishoeck2008}
{van Dishoeck} E.~F.,  {Jonkheid} B.,    {van Hemert} M.~C.,  2008, Fararaday
  Discussion of the Chemical Society, 133, 231

\bibitem[\protect\citeauthoryear{{Vander Wal}, {Scott} \& {Crim}}{{Vander Wal}
  et~al.}{1990}]{VanderWall1990}
{Vander Wal} R.~L.,  {Scott} J.~L.,    {Crim} F.~F.,  1990, J. Chem Phys., 92,
  803

\bibitem[\protect\citeauthoryear{{Vander Wal}, {Scott}, {Crim}, {Weide} \&
  R.}{{Vander Wal} et~al.}{1991}]{VanderWall1991}
{Vander Wal} R.~L.,  {Scott} J.~L.,  {Crim} F.~F.,  {Weide} K.,    R. S.,
  1991, J. Chem Phys., 94, 3548

\bibitem[\protect\citeauthoryear{{Visser}, {van Dishoeck}, {Doty} \&
  {Dullemond}}{{Visser} et~al.}{2009}]{Visser2009A&A...495..881V}
{Visser} R.,  {van Dishoeck} E.~F.,  {Doty} S.~D.,    {Dullemond} C.~P.,  2009,
  \aap, 495, 881

\bibitem[\protect\citeauthoryear{{Watson}}{{Watson}}{1976}]{Watson1976RvMP...4%
8..513W}
{Watson} W.~D.,  1976, Reviews of Modern Physics, 48, 513

\bibitem[\protect\citeauthoryear{{Willacy} \& {Woods}}{{Willacy} \&
  {Woods}}{2009}]{Willacy2009ApJ...703..479W}
{Willacy} K.,  {Woods} P.~M.,  2009, \apj, 703, 479

\bibitem[\protect\citeauthoryear{{Woitke}, {Kamp} \& {Thi}}{{Woitke}
  et~al.}{2009}]{Woitke2009A&A...501..383W}
{Woitke} P.,  {Kamp} I.,    {Thi} W.-F.,  2009, \aap, 501, 383

\bibitem[\protect\citeauthoryear{{Woitke}, {Thi}, {Kamp} \&
  {Hogerheijde}}{{Woitke} et~al.}{2009}]{Woitke2009A&A...501L...5W}
{Woitke} P.,  {Thi} W.-F.,  {Kamp} I.,    {Hogerheijde} M.~R.,  2009, \aap,
  501, L5

\bibitem[\protect\citeauthoryear{{Woodall}, {Ag{\'u}ndez}, {Markwick-Kemper} \&
  {Millar}}{{Woodall} et~al.}{2007}]{Wooddall2007A&A...466.1197W}
{Woodall} J.,  {Ag{\'u}ndez} M.,  {Markwick-Kemper} A.~J.,    {Millar} T.~J.,
  2007, \aap, 466, 1197

\bibitem[\protect\citeauthoryear{{Yung}, {Wen}, {Pinto}, {Pierce} \&
  {Allen}}{{Yung} et~al.}{1988}]{Yung1988Icar...76..146Y}
{Yung} Y.~L.,  {Wen} J.-S.,  {Pinto} J.~P.,  {Pierce} K.~K.,    {Allen} M.,
  1988, Icarus, 76, 146

\bibitem[\protect\citeauthoryear{{Zhang} \& {Imre}}{{Zhang} \&
  {Imre}}{1988}]{ZhangImre1988}
{Zhang} J.~Z.,  {Imre} D.~G.,  1988, Chem. Phys. Lett., 149, 233

\end{thebibliography}
\bsp

\label{lastpage}

\end{document}